\documentclass[journal=jacsat,manuscript=article]{achemso}
\setkeys{acs}{articletitle = true}
\usepackage[version=3]{mhchem} 
\usepackage[T1]{fontenc}       

\author{Leslie M. Schoop}
\affiliation{Department of Chemistry, Princeton University, Princeton, New Jersey 08544, USA}
\email{lschoop@princeton.edu}
\author{Florian Pielnhofer}
\affiliation{Max-Planck-Institut f\"ur Festk\"orperforschung, Heisenbergstra\ss e 1, D-70569 Stuttgart, Germany}
\author{Bettina V. Lotsch}
\affiliation{Max-Planck-Institut f\"ur Festk\"orperforschung, Heisenbergstra\ss e 1, D-70569 Stuttgart, Germany}
\alsoaffiliation{Ludwig Maximilian University, Munich 81377, Germany}
\alsoaffiliation{Nanosystems Initiative Munich \& Center for Nanoscience, 80799 Munich, Germany}
\email{b.lotsch@fkf.mpg.de}

\title{Chemical Principles of Topological Semimetals}

\abbreviations{2D, 3D, TI, DSM, WSM, NLSM, NSSM,...}

\begin{document}

\section{Preface}
The recent rapid development in the field of topological materials raises expectations that these materials might allow solving a large variety of current challenges in condensed matter science, ranging from applications in quantum computing, to infra-red sensors or heterogenous catalysis \cite{geim2009graphene,hosur2013recent,turner2013beyond,yang2011quantum,vsmejkal2017route,chan2017photocurrents,rajamathi2017weyl,habe2017tunneling}. In addition, exciting predictions of completely new physical phenomena that could arise in topological materials drive the interest in these compounds. \cite{son2013chiral,bradlyn2016beyond} For example, charge carriers might behave completely different from what we expect from the current laws of physics if they travel through topologically non-trivial systems \cite{xiong_evidence_2015,gooth2017experimental}. This happens because charge carriers in topological materials can be different from the normal type of fermions we know, which in turn affects the transport properties of the material. It has also been proposed that we could even find "new fermions", i.e. fermions that are different from the types we currently know in condensed matter systems as well as in particle physics \cite{bradlyn2016beyond}. Such proposals connect the fields of high energy or particle physics, whose goal it is to understand the universe and all the particles it is composed of, with condensed matter physics, where the same type, or even additional types, of particles can be found as so-called quasi-particles, meaning that the charge carriers behave in a similar way as it would be expected from a certain particle existing in free space. 

The field of topology in condensed matter physics evolved from the idea that there can be insulators whose band structure is fundamentally different (i.e. has a different topology) from that of the common insulators we know. If two insulators with different topologies are brought into contact, electrons that have no mass and cannot be back scattered are supposed to appear at the interface. These edge states also appear if a topological insulator (TI) is in contact with air, a trivial insulator. 2D TIs have conducting edge states whereas 3D TIs, which were discovered later, have conducting surface states. TIs have already been reviewed multiple times \cite{cava2013crystal,muchler2012topological,ando2013topological,yan2012topological}, which is why we focus here on the newer kind of topological materials, namely topological semimetals (TSMs). Nevertheless, we will refer to TIs and their properties where relevant in the context of topological materials and to contrast them with TSMs.

\begin{figure*}[htbp]
  \centering
  \includegraphics[width = \textwidth]{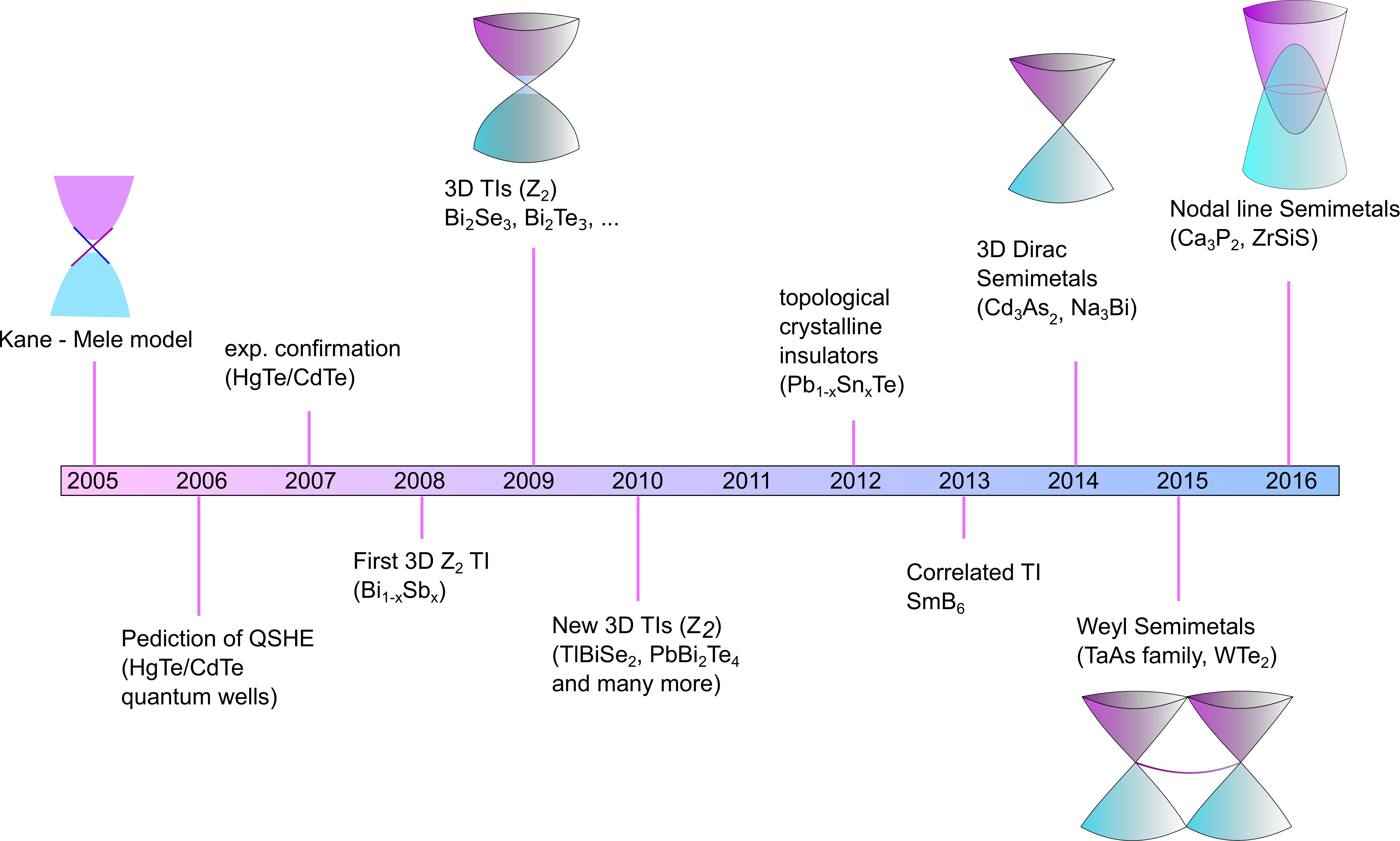}
  \caption{Timeline of the recent development in the field of topologically non-trivial materials.}
  \label{fig:FigT}
\end{figure*}

The term "topological semimetal" is widely used and basically includes all semimetals that exhibit some non-trivial band topology. We will describe in more detail later in this review what non-trivial topology actually means. Since TSMs are \textit{semimetals} they are characterized by a gap-less band structure, i.e. they differ from normal metals by being charge balanced, meaning the amount of electrons and holes contributing to their conductivity is exactly the same. Or, phrased differently, the hole and electron pockets composing the Fermi surface are of the same size. 

The number of materials predicted to be TSMs is extremely large \cite{yu2015topological, xie2015new, xu2017topological, wang2017new, yamakage2015line, feng2017topological, hirayama2017topological, schoop2015dirac, gibson2015three, seibel2015gold, weng2014transition, wan2011topological, zeng2015topological, yang2017topological, ruan2016ideal, liu2017nonmagnetic, chang2016strongly, chang2017type, du2017cate, zhu2016topological, chang2016kramers, chang2016room, xu2011chern}, however most of the experiments have only been performed on a handful of materials. This is usually linked to their chemical stability. Materials that can be grown easily and are additionally stable in environmental conditions are preferable to work with, and naturally, studied more often \cite{borisenko2014experimental,neupane2014observation,moll2016transport,li2015negative,ali2014crystal,schoop2015dirac,neupane2016observation,lv2016extremely,ali2016butterfly,hu2016evidence,emmanouilidou2017magnetic,wang2016body} Additionally, if we ever want to make topological semimetals technologically relevant or even scale them up, stability and ease of fabrication is of crucial importance, while the earth abundance, and hence, cost, of the materials should be kept in mind as well.

Chemists have a natural feeling for the stability, cost efficiency as well as level of difficulty to synthesize materials. Therefore, chemists should be able to identify the next exciting material in this field that will potentially lead the way to new types of quantum systems and electronics. However, in order to understand whether a material is a TSM, a significant amount of physical knowledge is required. Here, we will give an introduction to the necessary physics by using chemical principles, in addition to reviewing the known, experimentally verified TSMs. We will also explain the reason for the exotic behavior based largely on chemical principles, aiming to give chemists a guide for predicting and synthesizing new TSMs, and especially TSMs that are better candidates for future applications. 

This review is structured as follows: We will start with introducing the most famous TSM, graphene, from a chemical perspective. This includes a bit of history of the development of topological materials in general. We will then proceed by conceptually explaining the hallmarks of the different types of TSMs that have been reported to date, i.e. Dirac semimetals, Weyl semimetals, Dirac line node semimetals and nonsymmorphic semimetals. In the subsequent section we will review the most important experimentally verified TSMs, and finally we will briefly discuss some potential applications for such materials - predicted ones, as well as first experimental efforts toward these goals.

\section{From Graphene to Dirac Semimetals and Topological Insulators}

Graphene is famous for many reasons, particularly for it being the first example of an atomically thin material that could exist in a free standing way, i.e. without being supported by a substrate \cite{geim2007rise}. Besides its mechanical properties, it features exotic electronic properties \cite{novoselov2005two}. The basis of graphene's exotic electronic structure are two linearly dispersed bands that cross and form the so-called Dirac cone. This feature, as we will see, originates in carbon's special capability of forming delocalized, conjugated $\pi$-bonds. Additionally, symmetry, i.e. graphene's honeycomb structure, plays a crucial role. Models based on a honeycomb arrangement of atoms have fascinated theoretical physicists for a much longer time than the existence of graphene was confirmed; graphene's electronic structure was predicted by \textit{Wallace} as early as 1947 \cite{wallace1947band}. \textit{Duncan Haldane}'s toy model from 1988 \cite{haldane1988model} that was the basis of him winning the Nobel Prize of physics in 2016 \cite{Nobel2016}, was based on exactly such an arrangement. Much later, in 2005, \textit{Kane} and \textit{Mele} extended this model to graphene, thus starting the field of topology, which has been growing immensely fast ever since \cite{kane2005quantum}. 

\begin{figure*}[htbp]
  \centering
  \includegraphics[width = \textwidth]{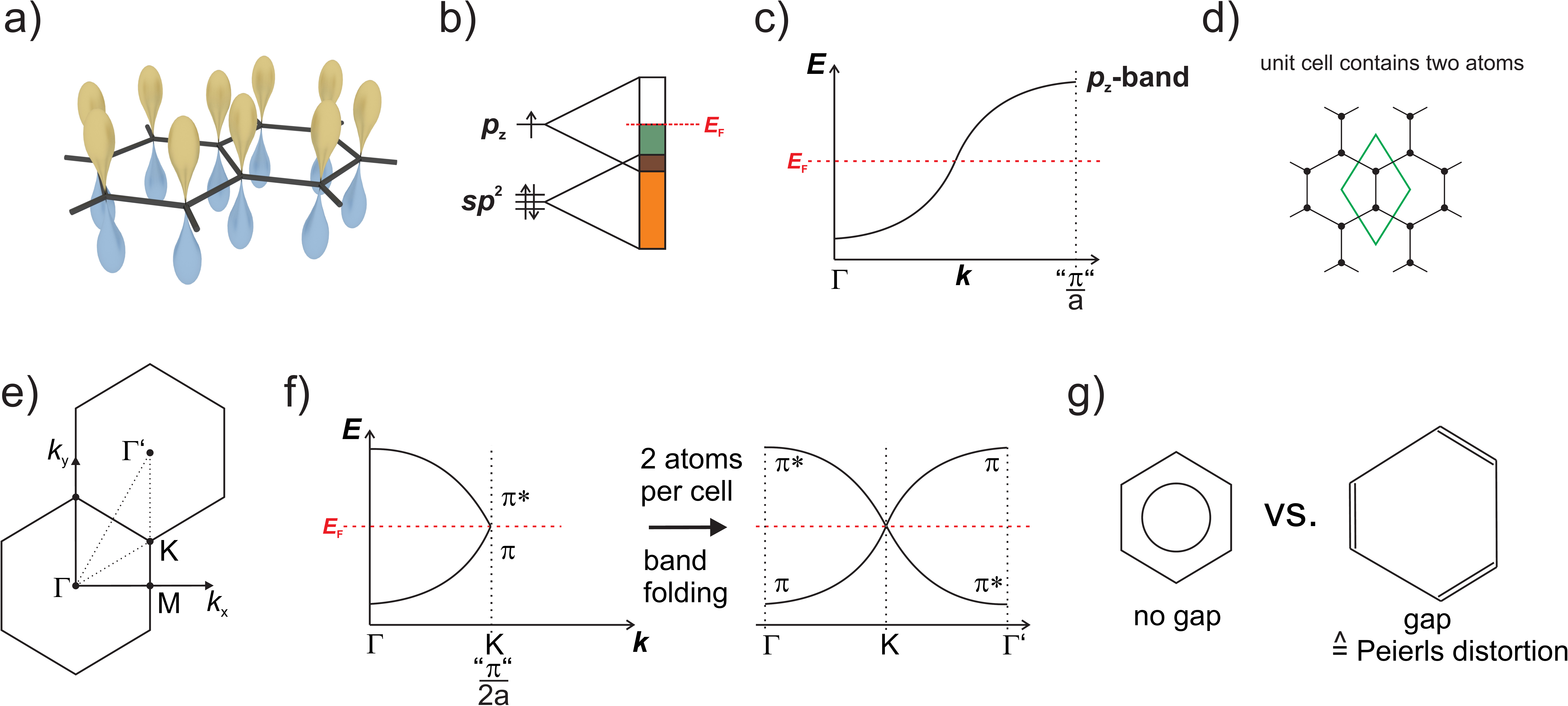}
  \caption{Intuitive approach for describing the electronic structure of graphene. (a) Real space structure of graphene, highlighting the delocalized $\pi$ system. (b) Orbital structure and band filling in graphene. (c) Corresponding electronic structure in \textit{k}-space; only one atom per unit cell is considered. (d) Unit cell of graphene, containing two atoms. (e) Brillouin zone of graphene. (f) Folded band structure of panel (c), in accordance with the doubling of the unit cell. (g) Real structure of graphene compared with a hypothetical, distorted version based on localized single and double bonds.}
  \label{fig:Fig1}
\end{figure*}

But lets first take a step back and look with a chemist's eyes at graphene and try to understand why it is so special. As chemists we would think of graphene as an \textit{sp$^2$}-hybridized network of carbon atoms. Thus three out of C's four electrons are used to form the $\sigma$-bonds of the in-plane \textit{sp$^2$} hybridized carbon backbone (Fig.\,\ref{fig:Fig1}a). The remaining electron will occupy the \textit{p$_z$}-orbital, and since all C-C bonds in graphene have the same length, we know that these electrons are delocalized over the complete graphene sheet. Since graphene is an extended crystalline solid, the \textit{p$_z$}-orbitals are better described as a \textit{p$_z$} band (Fig.\,\ref{fig:Fig1}b). Since there is one electron per C available, this \textit{p$_z$}-band is exactly half-filled (Fig.\,\ref{fig:Fig1}c). This, for a chemist intuitive description of graphene with a half-filled band, explains its metallic conductivity but not the presence of a Dirac cone in its electronic structure. Nor does a half-filled band \textit{per se} explain the unusual transport properties observed in graphene, although half-filling is a very rarely found scenario, an aspect we will come back to later. The reason for this model failing to predict the Dirac crossing in graphene is that we considered only a single C atom as a basis. The unit cell of graphene however contains two C atoms, due to symmetry reasons (Fig.\,\ref{fig:Fig1}d). Since band structures are plotted vs. the wave vector \textit{k}, which is inversely related to the size of the unit cell, doubling of the unit cell will bisect the \textit{k}-path (Fig.\,\ref{fig:Fig1}f). This scenario is very similar to \textit{Roald Hoffmann}'s famous description of a Peierls distortion in a hypothetical 1D chain of H atoms, and the reader is referred to this excellent paper for a deeper discussion of band foldings \cite{hoffmann1987chemistry}. 

We can thus basically view graphene as a folded band structure, analogous to Hoffmann's 1D chain of H atoms. But, since graphene is a hexagonal 2D material and not a 1D chain, we cannot draw the phases of the orbitals as easily as it is possible for a chain of H atoms. 
However, a simple tight binding model (see references \cite{wallace1947band, TB1, TB2, TB3}) shows that the bonding and the anti-bonding part of the \textit{p$_z$} band have to be degenerate at the K point. Thus, the simple intuitive way of folding the band structure for a unit cell containing more than one atom can be used to understand the basic electronic structure of graphene. For a 1D chain of H atoms, we would expect the next step to be a Peierls distortion that alternately forms a stronger bond between two neighboring atoms and thus, opens a band gap between the two degenerate states at the K point. Such a distortion, which would result in different bond lengths, is not observed for graphene. In chemical terms, the distortion is prevented by the delocalized $\pi$ system that stabilizes the perfectly hexagonal geometry energetically. A physicist would call the band degeneracy and resulting Dirac crossing "symmetry protected", since the hexagonal geometry guarantees that all bond lengths are the same, and thus, the bands have to be degenerate at K. Physicists and chemists thus view this issue from a different perspective, but come to the same conclusion. 

There are several ways the band degeneracy at K could be lifted in a honeycomb system, resulting in a band gap: (i) There could be a distortion, i.e. the bands would localize in double bonds, and the $\pi$-system would no longer be delocalized. This would open a trivial band gap and turn graphene into a trivial insulator. (ii) The electrons could be localized more at one atom than at the other in binary honeycomb systems such as for example in hexagonal 2D-BN sheets \cite{rubio1994theory}, where the electrons prefer to be located at the more electronegative nitrogen atoms. This also opens up a trivial gap, just as in (i). (iii) If a fourth bond is formed, as for example in graphite fluoride \cite{touhara1987structure}, or in hydrogenated graphane \cite{pumera2013graphane}, the C atoms will be \textit{sp}$^3$-hybridized and the delocalized $\pi$-system is destroyed, thus again, a trivial gap will open (with a few exceptions, see section 2D Materials below). (iv) The fourth possibility would be to add relativistic effects, i.e. spin orbit coupling (SOC), which increases the amount of mixing between orbitals. Thus, orbitals that were orthogonal to each other and would not mix without SOC, will mix if there is large SOC, thus creating a band gap. This opens a gap in a non-trivial way and creates a 2D TI, as discussed in the 2D Materials section below. Note that for this to happen, the $\pi$ system is still partly intact and not localized or fully \textit{sp}$^3$-hybridized. This scenario was the basis of the \textit{Kane-Mele} model in 2005, the ground work for the prediction and later discovery of topological insulators  \cite{kane2005quantum}. Technically, even in graphene, SOC should open up a very small band gap, which is too small to be observed experimentally \cite{gmitra2009band}. So why then are the most known TIs not based on delocalized $\pi$ systems? It is obvious that the higher homologs of graphene and, hence, its high SOC analogs, are no stable compounds, as $\pi$ bonds become inherently less stable if we go down the periodic table.
Thus, although many theoretical predictions of exciting properties in hypothetical planar 2D sheets of silicene, germanene or stanene appeared in high impact journals \cite{liu2011quantum, ni2011tunable,cahangirov2009two}, their realization is extremely difficult. These materials will want to form a fourth sigma bond, and thus buckle to ultimately become 3D Si (Ge, Sn). \textit{Roald Hoffman} describes this problem nicely in his "silicene interlude" incorporated in a statement about nanoscience \cite{hoffmann2013small}. We will discuss the experimental advances of such 2D TIs and TSMs more in section "2D Materials" later. 

Taking all known materials that are TIs into account, the TIs that are not based on a conjugated $\pi$ system dominate, in fact. The first prediction for this class of TIs was made by \textit{Bernevig}, \textit{Hughes} and \textit{Zhang} (the so-called BHZ model) in 2006 \cite{bernevig2006quantum}. In their original paper they introduce the possibility of a band inversion causing topological edge states. The closest analogy of a band inversion to chemical principles observed in molecular systems might be complexes that contain an inverted ligand field. It has been shown that it is indeed possible that molecules can have ligand field splittings that are the opposite to what is expected \cite{snyder1995elusiveness,jenkins2002elucidation}, i.e. 3 over 2 in an octahedral environment. For an excellent review about inverted ligand fields, please see \cite{hoffmann2016widely}. Similarly, bands can also be inverted, meaning that the order in energy is different from what is expected. \textit{Bernevig}, \textit{Hughes} and \textit{Zhang} used this idea to predict HgTe to be a 2D TI if made thin enough, which was experimentally confirmed soon after by \textit{Molenkamp et al.} \cite{konig2007quantum}. Due to the lanthanide contraction and the resulting inert pair effect, the \textit{6s}-orbital in mercury is particularly low in energy, thus it can be below the Te \textit{5p}-states and the bands are inverted at the $\Gamma$ point. The inert pair effect is not the only possibility to have a band inversion, but many of the known topological insulators are based on heavy elements such as Bi or Pb, thus using this effect. The important difference between a band inversion in a solid and a molecule is that a solid can have an inverted band structure at one location in the Brillouin zone (BZ), but will be "normal" in other parts. This is the crucial requirement for it being topological. Since the BZ of a molecule only consists of the $\Gamma$ point, it can either be inverted or not, and it is not possible to switch between an inverted and un-inverted electronic structure, which is why molecules cannot be TIs.

Note that also graphene has an inverted band structure, just for a different reason than HgTe. Due to the band folding the $\pi^\ast$ band is below the $\pi$ bands in the second BZ. This can be seen in Fig.\,\ref{fig:Fig1}f (right panel), where the band structure is plotted not only in the first, but also the second BZ. Thus graphene's band inversion is within the delocalized $\pi$-system. The inverted band structure is indeed the reason that SOC would create a non-trivial gap in heavy graphene. Thus band inversions are in general a good thing to look for, if trying to find new topological materials. Chemists usually have an intuitive understanding of when the order of the orbitals might be mixed up and consequently, a band inversion is likely to appear. From molecules we know for example that the combination of a quite electronegative metal and an electropositive ligand can invert the orbital order (i.e. invert the ligand field, see above), since a higher fraction of the bonding electrons can be attributed to the metal \cite{hoffmann2016widely}. Thus chemical intuition can help in the search of new topological materials.

\section{Types of topological semimetals}

Most TSMs have in common that their unusual band topology arises from a band inversion. Unlike TIs, they do not have a band gap in their electronic structure. There are several classes of TSMs: Dirac semimetals (DSMs), Weyl semimetals (WSMs) and nodal line semimetals (NLSMs). All these kinds exist as "conventional" types, i.e. they are based on a band inversion. In addition, they can also be protected by nonsymmorphic symmetry. The latter ones have to be viewed differently and we will discuss them after introducing the conventional ones.

\subsection{Dirac Semimetals}

\begin{figure*}[htbp]
  \centering
  \includegraphics[width = 0.85\textwidth]{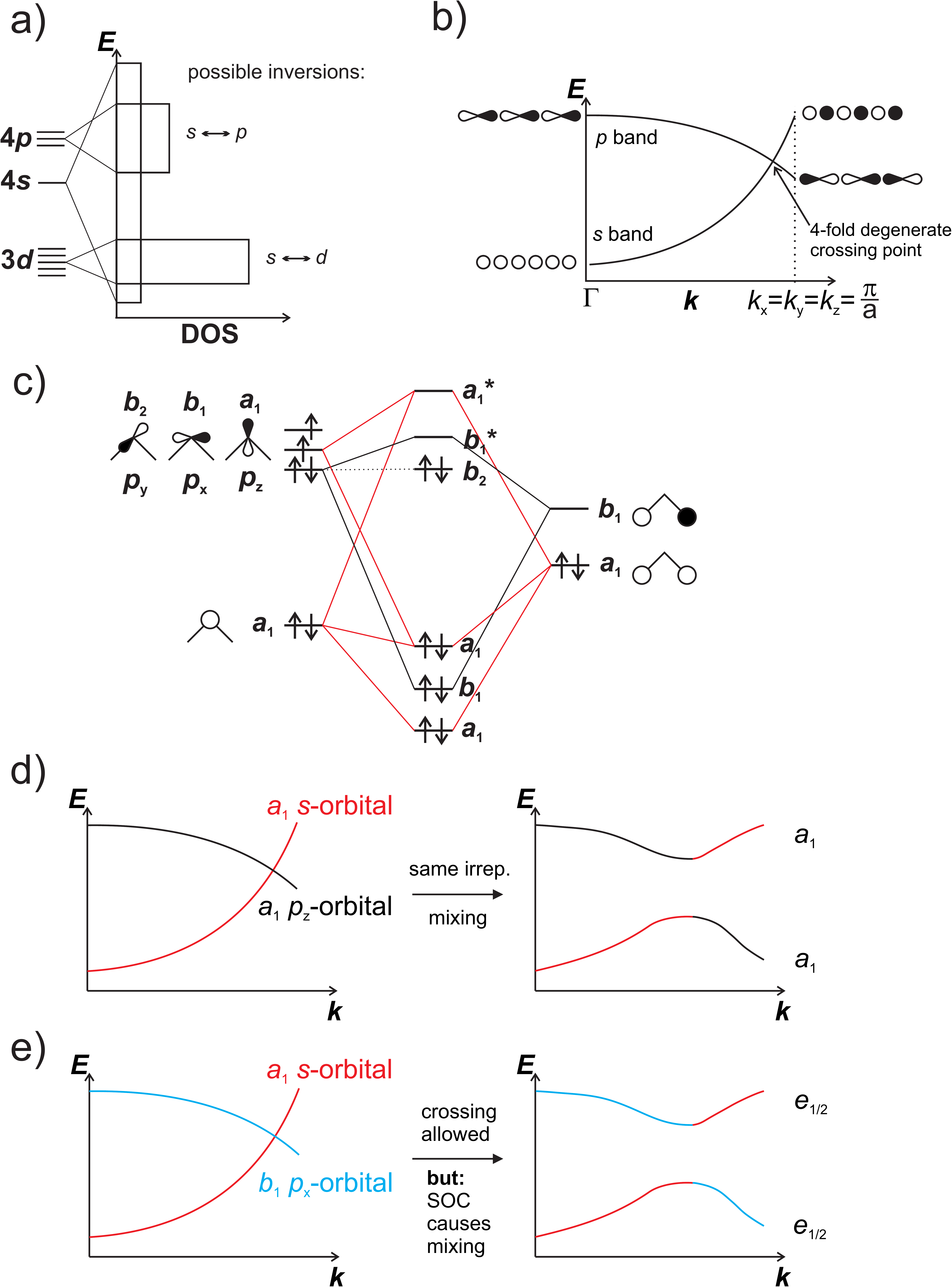}
  \caption{Explanation of band inversions. (a) Rough density of states (DOS) of transition metals. Band inversions are possible between the different orbitals within one shell, but the material is likely to be metallic. (b) Band inversion between a \textit{s} and a \textit{p} band. (c) MO diagram of water. (d) Bands that cross and have the same irreducible representation (irrep) gap. (e) If the irreducible representations are different, the crossing is protected, but SOC might still create a gap.}
  \label{fig:Fig2}
\end{figure*}

The prototype of a Dirac semimetal is graphene. The "perfect" DSM has the same electronic structure of graphene, i.e. it should consist of two sets of linearly dispersed bands that cross each other at a single point. Ideally, no other states should interfere at the Fermi level. Note that in a DSM, the bands that cross are spin degenerate, meaning that we would call them two-fold degenerate and thus the Dirac point is four-fold degenerate. When discussing degeneracies within this review, we will always refer to spin-orbitals. In any crystal that is inversion symmetric and non-magnetic (i.e. time reversal symmetry is present) all bands will always be two-fold degenerate. Thus, if two bands cross in an inversion-symmetric, non-magnetic material, there will be a Dirac crossing, as long as the crossing is stable (see Fig.\,\ref{fig:Fig2}a-b). The question we have to address is thus when such a crossing can appear and whether the crossing is stable. When two bands cross, it means that they are inverted after that crossing point. Generally that is not really hard to achieve in a solid, because bands have a band width, and are thus likely to overlap as they move away from the $\Gamma$ point. However, this happens most commonly in open-shell systems, i.e. compounds that are not charge balanced. In such systems, it will be very hard to find a sole Dirac cone at the Fermi level, with no other states crossing. 

Thus, to realize charge balanced systems where bands are nonethlelss crossing within the BZ, we are looking for closed shell systems, where the energy of the bands of the closed shell overlap with the ones of the next, empty shell. Ideally, just two bands cross at one point in the BZ. This means we are looking for materials where the bonding and/or the electronegativity difference is neither too weak nor too strong, because the latter will induce a large gap between filled and empty states. 

\clearpage

\clearpage

However, the next complication arises from the fact that not every band crossing that appears due to overlapping shells is stable. Crossings can be avoided, which basically means that a gap is created at the crossing point. There is a simple rule to understand whether a crossing is allowed or not; it depends on the orbital (band) symmetry and whether or not mixing between the orbitals (bands) is allowed. Take for example the molecular orbital (MO) diagram of water. All three \textit{a$_{1g}$} orbitals - one arising from the ligand group orbitals formed by the two H's and the others from the O-\textit{2s}-and \textit{p$_z$}-orbitals, respectively (see Fig.\,\ref{fig:Fig2}c) - are allowed to mix as they are assigned the same irreducible representation in the \textit{C}$_{2v}$ character table.  


Similarly, if two bands that have the same irreducible representation cross, they can mix and the crossing will be avoided and thus a gap will open (Fig.\,\ref{fig:Fig2}a-d). If the two crossing bands have different irreducible representations, the crossing will be allowed (allowed to mix is hence the opposite of allowed to cross). Thus, in graphene, the $\pi$- and $\pi^\ast$-bands must have different irreducible representations, which is indeed the case (in the absence of SOC) \cite{kogan2012symmetry}. The point group along the $\Gamma$-K line in graphene is \textit{C}$_{2v}$ and the $\pi$- and $\pi^\ast$-bands can be assigned to the irreducible representations $a_2$ and $b_2$, respectively. At the K point, which has \textit{D}$_{3h}$ symmetry, they merge to a doubly degenerate $e^{\prime\prime}$ irreducible representation\cite{kogan2012symmetry}. 
But why would SOC affect a degeneracy?
In the presence of a significant amount of SOC, point groups have to be extended to (electron spin) double groups, which account for half integer angular momentum states \cite{heine2007group}. Double groups add more irreducible representations to a point group, but only the newly added ones are accessible in the presence of high SOC, thus the effective number of irreducible representations decreases. For example, in the presence of SOC, the \textit{C}$_{2v}$ double group contains only a single irreducible representation (\textit{e}$_{1/2}$), meaning that all orbitals are allowed to mix in a heavy \textit{C}$_{2v}$ molecule. We can also think about SOC as increasing the likelihood of orbital mixing. So we can look at all 32 point groups, extended to double groups, and figure out when, in the presence of SOC, crossings are possible. If we do this, we will see that this is only the case if we have either a \textit{C}$_3$, \textit{C}$_4$ or \textit{C}$_6$ symmetry element (see also ref. \cite{gibson2015three} for a discussion of this matter), which is why we expect to find 3D DSMs only in highly symmetric systems. When we determine irreducible representations of bands, we need to look at the local point group of where in the BZ the crossing is appearing. Thus, if we are away from a high symmetry line or point, the point group will be of low symmetry and consequently, we can expect that most crossings are avoided, especially in the presence of SOC. In the case of graphene, a 2D material, the BZ is only defined for directions within the 2D sheets, thus there is no possibility to have an in-plane allowed crossing on a high symmetry line in the presence of SOC in 2D, since the maximal rotational axes present in the local point groups of the high symmetry lines is \textit{C}$_2$. In 3D, we have more options, which led to the development of 3D Dirac semimetals. 3D DSMs have four-fold degenerate points in their electronic structure that are robust against SOC.

Let us briefly comment on SOC-induced gap opening. If a crossing is avoided and gapped by SOC, this does not necessarily mean a compound is trivial, since there can still be the band inversion. Just as in the case of heavy graphene, this could result in a TI and thus, in conducting edge states that will also be promising for device applications. Note, however, that the material will not necessarily be a TI; it can also be trivial, depending on the total number of band inversions, which have to be odd in the case of TIs.

What interesting physics is to be expected from 3D DSMs? For one, we expect that the resulting Dirac electrons have a very low mass (ideally 0) which will result in them being extremely mobile and thus these materials are very conductive, depending on their charge carrier concentration. Dirac fermions basically behave more like photons rather than electrons, which intuitively can explain why they are so fast. We will discuss the properties of Dirac semimetals more in detail below. Another reason for the interest in DSMs is that if the bands are spin split, they turn into Weyl semimetals (WSMs). These host the Weyl fermion, which is a fermion that is chiral and different from the usual Dirac fermions we know. Although this fermion was predicted to exist in 1929 \cite{weyl1929elektron}, it was only discovered in 2015 \cite{weng2015weyl,xu2015discovery,lv2015experimental,xu2015discovery2, yang2015weyl,lv2015observation}, which was a big discovery for both, condensed matter physics and particle physics.

\subsection{Weyl semimetals}

The difference between a Dirac- and a Weyl semimetal is that in the latter, the crossing point is only two-fold degenerate \cite{wan2011topological,balents2011weyl,vafek2014dirac}. This is because in WSMs, the bands are spin split, thus each band is only singly degenerate. If a spin-up and a spin-down band cross, this results in a Weyl monopole, meaning that there is a chirality assigned to this crossing. Since there cannot be a net chirality in the crystals, Weyl cones always come in pairs. The resulting Weyl fermions are chiral in nature and thus will behave physically different from "regular" fermions. One example of this manifestation is the chiral anomaly, which we will discuss in the properties section below. Here, we will focus on the requirements necessary to realize a WSM.

\begin{figure*}[htbp]
  \centering
  \includegraphics[width = \textwidth]{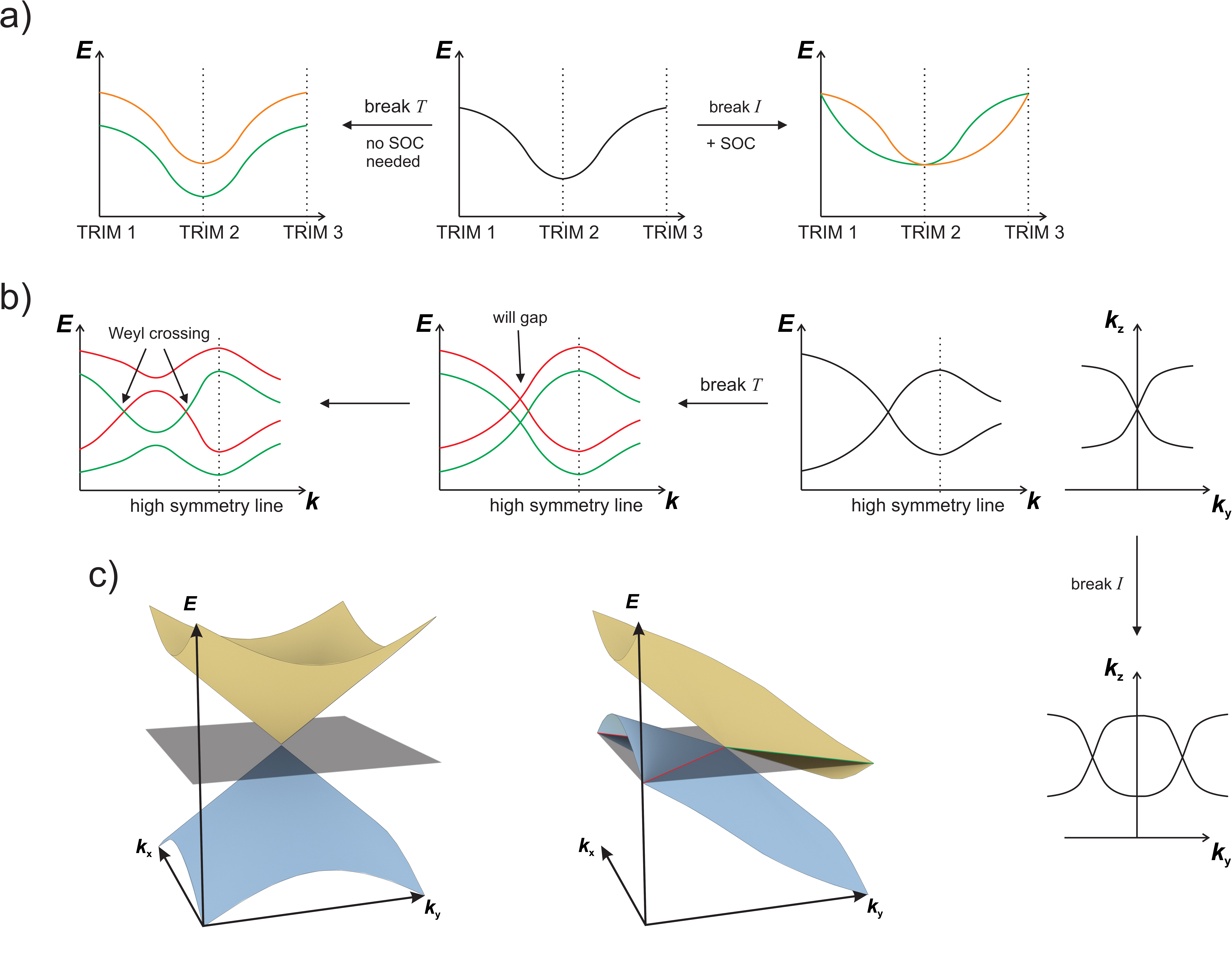}
  \caption{The different ways to achieve a WSM. (a) Effect of \textit{T}- and \textit{I}-symmetry breaking on a single band. (b) The same scenario for a Dirac crossing. In the case of \textit{T} breaking, two Weyl crossings will appear on the high symmetry line at different energies. In the case of \textit{I} breaking, they will appear away from the high symmetry line. (c) Schematic drawing of a type I (left) and a type II (right) WSM.}
  \label{fig:Fig3}
\end{figure*}

In order to have spin split bands, we cannot have inversion (\textit{I}) \textit{and} time-reversal (\textit{T}) symmetry at the same time, since the combination of these two symmetries will always force all bands to be doubly degenerate. In \textit{I} asymmetric, i.e. non-centrosymmetric crystals, this degeneracy can be lifted with the help of SOC; this is the so-called Dresselhaus effect \cite{dresselhaus1955spin}. This effect only appears away from the time-reversal inversion momenta (TRIMs), which are high symmetry points in the BZ (only the ones where \textit{k}$_x$,\textit{k}$_y$ and \textit{k}$_z$ are all either 0 or 0.5) (see Fig.\,\ref{fig:Fig3}a). At the TRIM points, \textit{T}-symmetry still enforces two-fold degeneracy. Thus, in inversion asymmetric compounds, Weyl crossings appear away from high symmetry lines. Since \textit{T}-symmetry is still present, the Weyl cones will be multiplied to a minimum number of four. If further crystal symmetry acts on them (i.e. rotates or reflects them), there can be many more Weyl cones, for example there are 24 in TaAs and related materials \cite{weng2015weyl, sun2015topological}. In order to have a clean system to carefully study the properties of Weyl fermions, it would be highly preferred to have as few Weyl cones as possible in the BZ. The second possibility for creating a WSM would be to try to loose \textit{T}-symmetry \cite{wan2011topological} (Fig.\,\ref{fig:Fig3}a). This can be achieved easily by applying a magnetic field, however the effected (Zeeman) band splitting is only relatively small, even for very high fields. Much stronger splitting can be observed in intrinsically magnetic samples. The magnetic order needs to be either ferro- or ferrimagnetic, or some canted version of antiferromagnetism to sufficiently break \textit{T}-symmetry. There are not yet many examples of magnetic WSMs. In order to predict a magnetic WSM reliably, the exact magnetic structure should be known. But so far, the amount of materials with published magnetic structures is small compared to the total number of known materials, which slows down the search tremendously. \textit{T}-symmetry breaking splits the bands with a different mechanism than \textit{I}-symmetry breaking (see Fig.\,\ref{fig:Fig3}a), allowing for Weyl cones to be on high-symmetry lines (see Fig.\,\ref{fig:Fig3}b). Therefore, it is possible to have just two Weyl cones in the BZ in a magnetic semimetal (the lowest possible number due to chirality), making these types of WSMs generally superior to the inversion asymmetric ones. A material that just has two Weyl cones, the so-called "hydrogen-atom" of the WSMs, has not yet been found.

It is harder to predict a WSM with chemical principles. Since the crossings are away from any high symmetry regions in inversion asymmetric semimetals, the search is very much obscured. In the case of time-reversal breaking WSMs, chemists can however be very helpful. We know which elements in a material are likely to induce ordered magnetism and can even, with the help of some rules (i.e. the Goodenough-Kanamori rules\cite{anderson1950antiferromagnetism, anderson1959new, goodenough1955theory, goodenough1958interpretation, kanamori1959superexchange}, or the Bethe-Slater curve \cite{slater1930cohesion, sommerfeld1933handbuch, bozorth1951ferromagnetism}) guess what type of magnetic coupling we can expect. That will allow us to look at materials that are known to exist, but may have magnetic properties that are unknown. 

In addition to regular type I WSMs, the existence of so-called type II WSMs \cite{soluyanov2015type} has been confirmed in layered transition metal dichalcogenides \cite{tamai2016fermi,deng2016experimental}. Type II Weyl semimetals also have twofold degenerate band crossings, but the cones are tilted (see Fig.\,\ref{fig:Fig3}c) so that it is not possible to have a zero density of states at the Fermi level; there will always be electron and hole pockets. The Weyl point appears exactly at the contact between the electron and hole pockets. Type II WSMs have been proposed to behave electronically different form type I WSMs \cite{soluyanov2015type}.

\subsection{Nodal line Semimetals}

There are examples of Dirac and Weyl semimetals, where the four- or twofold degeneracy is not just a point, but a line or a circle \cite{burkov2011topological,fang2015topological,fang2016topological,PhysRevLett.115.036806,xie2015new}. In contrast to a normal Dirac or (type I) Weyl semimetal, where the Fermi surface should just be a point (or several points), nodal line semimetals (NLSMs) have loops or lines as Fermi surfaces. Thus, the concentration of Dirac or Weyl electrons is much higher in these materials, which makes them preferable candidates for certain types of electronics, such as magnetic sensors \cite{singha2017large}. NLSMs can also host exotic, extremely flat "drumhead" surface states that have been suggested to lead to superconductivity or magnetism at the surface of these materials \cite{chan2015topological,liu2017correlation}. For an overview of NLSMs, the reader is referred to a recent review article \cite{yang2017symmetry}. 

\begin{figure*}[htbp]
  \centering
  \includegraphics[width = \textwidth]{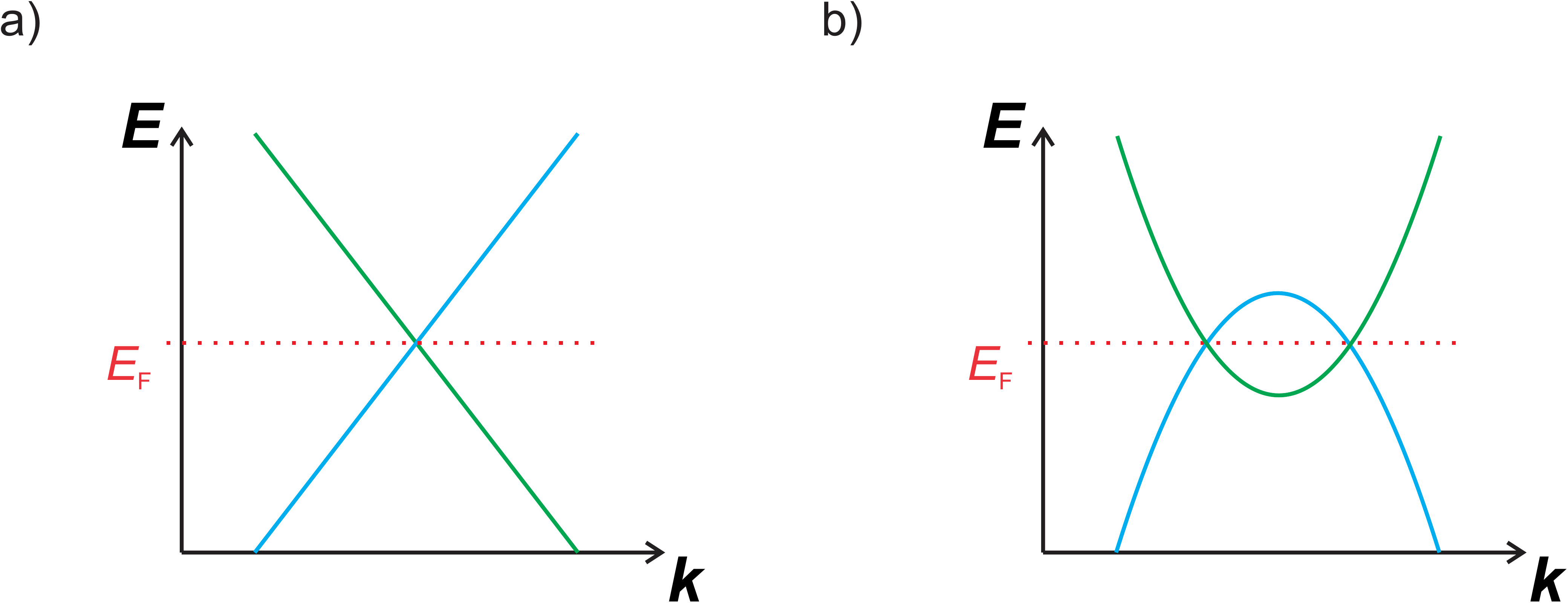}
  \caption{Schematic band structure of a) a Dirac cone and (b) a nodal line. While the Fermi surface of the former is only a point, the one of the latter is a loop.}
  \label{fig:Fig66}
\end{figure*}

In order to understand how to find materials that can be NLSMs, it is useful to divide them in two categories: four-fold degenerate Dirac NLSMs and two-fold degenerate Weyl NLSMs (see above). For the latter case there are only very few examples \cite{bian2016topological} in non-centrosymmetric compounds and no magnetic ones are known yet. There are however quite a few examples of Dirac NLSMs \cite{schoop2015dirac,takane2017observation,xie2015new,yu2015topological,feng2017topological}. They all have in common that they are low SOC materials. Since a crossing needs to be allowed over a full line or loop of \textit{k}-points in the BZ, SOC is very likely to gap it at lower symmetry points of this loop/line. The known NLSMs are usually protected by mirror planes, which means that the mirror plane is what allows the bands to have two different irreducible representations. This can only work without the presence of SOC, thus in order to find NLSMs we need to look for materials with light elements that still feature a band inversion. Compounds composed of relatively light elements that were classified as a NLSM are discussed in the section "Material examples" below.

\subsection{Nonsymmorphic topological semimetals}

There is a second, completely different way of achieving topological semimetals that is not necessarily based on a band inversion. Rather, this concept is based on nonsymmorphic symmetry elements. These are symmetries such as glide planes or screw axes, which have the characteristic that they include a translation. Thus they only appear in space groups, rather than point groups, since the spatial origin is not conserved. The extra translation symmetry element doubles (or further increases) the content of the unit cell. Therefore, we can again think of the band structure of nonsymmorphic materials as being folded. This time, the induced band degeneracies that appear at the BZ edges and corners will be forced to remain degenerate as long as the nonsymmorphic symmetry is preserved. This means they could technically be lifted by a structural distortion, but they are immune to gapping due to SOC. The concept of inducing band degeneracies at BZ edges in nonsymmorphic space groups has first been described by \textit{Michel} and \textit{Zak} in 1999 \cite{michel1999connectivity} and was used by \textit{Rappe et al.} to predict the first 3D Dirac semimetal in 2012 \cite{young2012dirac}. \textit{Young }and \textit{Kane} later expanded the concept to 2D materials, where nonsymmorphic symmetry gives the advantage that the crossing can be stable in 2D, since SOC shows no effect \cite{young2015dirac}. Finally, it was shown by \textit{Bernevig et al.} that the concept can also be used to induce three-, six- and eight-fold degeneracies \cite{bradlyn2016beyond} (the eight-fold ones were simultaneously characterized by \textit{Kane et al.}\cite{wieder2015double}), and that these new types of degeneracies can lead to exotic "new fermions". These fermions are supposed to behave differently in transport than Dirac or Weyl fermions, and in contrast to these, they do not have any counterparts in high energy physics. Thus, they can only exist in the solid state and it is a great challenge for solid state chemists to realize these new quasi-particles. It is also kind of intuitive why for example a three-fold band degeneracy can be special - in this case we have basically a Weyl fermion with a 3/2 spin. The three-fold degenerate new fermions are the most exciting ones of the potential new fermions, as they are predicted to show surface states that have not been seen yet and in addition, these fermions should move unexpectedly under the influence of a magnetic field \cite{bradlyn2016beyond}. Note that three-fold degeneracies can also appear accidentally in inversion-asymmetric crystals. There are a few examples\cite{weng2016topological, lv2017observation, shekhar2017extremely, winkler2016topological, zhu2016triple}, but in these materials a doubly degenerate band crosses a singly degenerate band, which does not result in a "new fermion".
It has been fully characterized what kind of band degeneracy can be expected at which high symmetry line or point in all space groups \cite{watanabe2016filling,watanabe2015filling}, and it has been tabulated in the Bilbao crystallographic server \cite{aroyo2006bilbao2,aroyo2006bilbao1,aroyo2011crystallography,elcoro2017double}. Therefore, we could technically just look at materials in space groups that are listed as being hosts for higher order degeneracies and where such crossing points \textit{have} to be present, as they are demanded by group theory. In contrast to "conventional" TSMs, the orbital character of the band does not influence the band degeneracy, but solely the symmetry of the space group and, hence, the presence of nonsymmorphic symmetry.

Sadly, the most common problem with achieving these proposals is that the Fermi energy is usually not at the crossing point.  Since the band structure was folded to accommodate more atoms in the unit cell, we would need a half-filled band (just like in graphene) for the Fermi level to be located at the crossing point (see also  \cite{gibson2015three,topp2016non,yang2017symmetry} for a discussion of this issue.) Half-filled bands, especially when isolated, are usually extremely unstable, with graphene being an exception due to its delocalized $\pi$ system. Half-filled \textit{d} bands are famous for distorting and forming charge density waves, or the electrons are so localized that the compounds become Mott insulators, in order to avoid half-filling. \cite{wilson1975charge,austin1970metallic,zaanen1985band} Half-filled \textit{s} or \textit{p} bands basically never exist; just like in the hypothetical 1D chain of H atoms, they will distort in order to form covalent bonds \cite{hoffmann1987chemistry}. This is nicely summarized by the Jahn-Teller theorem which states that a "nonlinear molecule with a spatially degenerate electronic ground state will undergo a geometrical distortion that removes that degeneracy, because the distortion lowers the overall energy of the species" \cite{jahn1937stability}. Like in molecules, this can be applied to solids if the Fermi level resides at a multiply degenerate band crossing point. Thus the electron filling in nonsymmorphic materials is usually such that the whole band is filled and the Fermi level lies in the trivial gap. If the Fermi level is not within a band gap, the compound is usually metallic and has enough Fermi surface pockets to stabilize a nonsymmorphic crossing relatively close to the Fermi level. In this case, the transport signature of the trivial pockets will obscure the signal of the exotic fermions. This problem has so far been the greatest challenge in observing the transport signatures of fermions that are the result of band degeneracies enforced by nonsymmorphic symmetry.

Ordered magnetism will have more of an effect on nonsymmorphic materials than "just" breaking time reversal symmetry. In addition to ferro- and ferrimagnetism, antiferromagnetism can also affect the band degeneracies because it introduces new symmetry elements. The spin order creates symmetry, and this symmetry can be either symmorphic or non-symmorphic. If magnetism is included, the 230 space groups extend to 1651 magnetic groups. First works on trying to classify all possible band degeneracies enforced by symmetry in magnetic groups have been published \cite{watanabe2017structure}, but not many of these potential "magnetic new fermions" have been explored yet. We were able to show recently that it is possible to use antiferromagnetic order in CeSbTe to merge two four-fold degenerate bands (each point being represented by a four-dimensional irreducible representation) to an eight-fold degenerate point, represented by an eight-dimensional irreducible representation \cite{schoop2017tunable}. The cause of this is the extra screw axis induced by the spin order. More information about CeSbTe can be found in the "Material Examples" section. To summarize, the field of magnetic nonsymmorphic materials is still young and will need much further investigations, but already now holds a lot of promises for the future.

\section{A Quantum Chemical Approach to Topology}

Recently a new description of topological materials was introduced by a group of theorists at Princeton University. It allows to predict topological materials just by considering the type of atomic orbitals located on a certain Wyckoff position in a known space group \cite{bradlyn2017topological}. For this, they classified all possible elementary band representations (EBRs) in all 230 space groups. An EBR is a classification of the symmetry or irreducible representation of an orbital if it is repeated in a crystallographic lattice. Let's for example consider a simple cubic lattice of hydrogen, where we just have one 1\textit{s}-orbital located on the \textit{1a} Wyckoff position (0,0,0) in space group \textit{Pm$\bar{3}$m}. These 1 \textit{s} atomic orbitals would generate one EBR. If we would consider \textit{p}-orbitals as well, all orbitals that have the same irreducible representation in real space (such as $a_1$ or $b_2$) would each form a separate EBR. 
We can also think of EBRs as the molecular orbitals in a solid. 
An EBR is thus a \textit{set} of bands, i.e. several bands can belong to a single EBR, just like several molecular orbitals can have the same irreducible representation (i.e. there are three $t_{2g}$ orbitals in octahedral transition metal complexes). Although an EBR is formed by the same type of orbitals in real space (i.e. \textit{s}, \textit{p} or \textit{d}), it can yield a sum of different irreducible representations at each point in \textit{k} space.  Since bands are extended in space, the irreducible representation of an EBR can change in \text{k}-space. One can imagine this if one considers that the symmetry at different \textit{k}-points can be different, i.e. the $\Gamma$ point is more symmetric than a \textit{k}-point on the $\Gamma$-X line. 

\begin{figure*}[htbp]
  \centering
  \includegraphics[width = \textwidth]{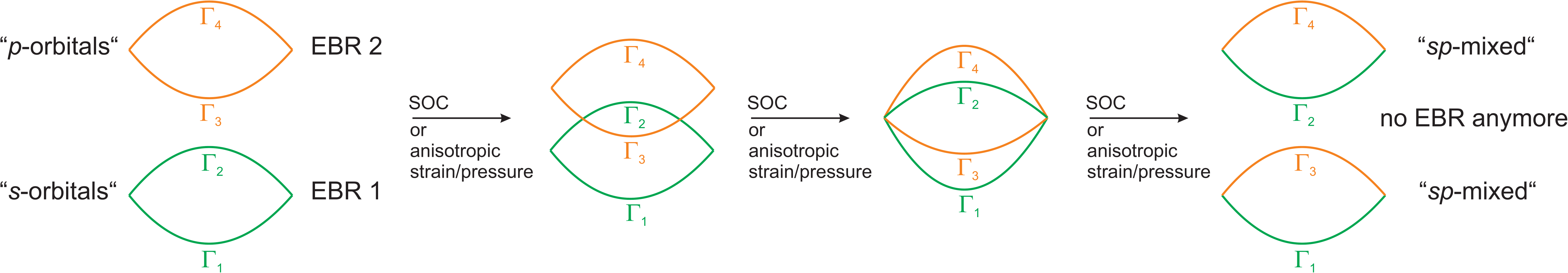}
  \caption{Schematic drawing of EBRs. Two EBRs mix to create a TI, the $\Gamma$ labels represent the irreducible representations.}
  \label{fig:Fig4}
\end{figure*}

All 10,403 possible EBRs have been fully characterized for all space groups and all maximal Wyckoff positions (which are the ones where there is no variable coordinate). The Bilbao crystallographic server can be used to obtain the EBRs for a material of interest \cite{vergniory2017graph,bradlyn2017band,elcoro2017double,cano2017building}.  But how can we use EBRs to classify if a material is topologically non-trivial? Let us assume we have an EBR that consists of four bands. It could be that all four bands can be connected at a certain \textit{k} point in the BZ; in that case the material will be trivial and the Bilbao crystallographic server will mention that this is a fully connected EBR. If however only two of the four bands are connected, then there is a possibility that there is a complete gap within a single EBR. In this case the EBR would be called disconnected, and we will have a topological insulator. Disconnected EBRs cannot exist in an atomic limit, which is analogous to the fact that molecules cannot be TIs, as explained earlier. The Fermi level also plays an important role. If the Fermi level lies within a single, connected EBR, the material will be a semimetal. Graphene is an example for this case. If now such a half-filled EBR gets disconnected (i.e. a gap opens), the material becomes a TI. This can happen with a "tuning knob" such as SOC or strain/pressure (with no accompanied symmetry change). TIs are therefore materials where a disconnectecd EBR is only partially filled but the material still has a band gap. There is a second possibility to create a TI, if two EBRs are involved. Two different EBRs can overlap and create a band inversion that way, as shown in Fig.\ref{fig:Fig4}. If the EBRs reopen after inverting to form a band gap, we have a "composite" band representation, where the two groups of bands together are a sum of two EBRs, but neither the bands above nor below the gap correspond to a single EBR (see Fig.\,\ref{fig:Fig4}). An example for this scenario is Bi$_2$Se$_3$.

Using the Bilbao crystallographic server, it is now possible to get a list of irreducible representations that can exist at high symmetry points. Then the EBRs can be built by connecting the irreducible representations by hand, in a way that is compatible with the group representations, just as we would connect the atomic orbitals with the same irreducible representations to molecular orbitals. Thus, with such a list of EBRs and pen and paper, we can guess if we have a topological material - before performing a DFT calculation.\\


\section{Material examples} \label{sec:Mater}

The experimental search for topologically non-trivial materials is usually driven by theoretical models that predict unusual electronic properties in crystalline solids. These models identify features such as band crossings, band inversions, Dirac cones, Weyl nodes, etc. as the fingerprints of topology in the electronic band structure of candidate materials. Thus, the obvious, but tedious way to find suitable candidate materials is by screening electronic band structures of known materials. Thinking within chemical reason can speed up the search, for example by looking for patterns in structural motifs. Structures based on honeycomb lattices like in graphene or square lattices as found in many DSMs can therefor be a good starting point for a chemist to develop ideas for synthesizing novel materials. As solid state chemistry is always related to crystallography and, for sure, electronic structure theory, chemists are capable of predicting and synthesizing new topologically non-trivial materials. Once a known compound is predicted, the experimental challenge of confirming the prediction by angle-resolved photoemission spectroscopy (ARPES) and transport measurements on appropriate crystals begins.
As we describe above, employing chemical intuition together with crystallographic concepts  opens a possibility for more chemists to join the field of topologically non-trivial materials.
After the first experimental confirmation of a topological semimetal \cite{liu2014discovery,borisenko2014experimental, liu2014stable, neupane2014observation} the number of further contenders, either predicted or verified, increased rapidly. Below we will provide a short overview of topological semimetals but restrict ourselves to examples that are experimentally verified and well-studied.

\subsection{3D Dirac semimetals}
The first Dirac semimetal to be experimentally verified was Na$_3$Bi, a compound that was synthesized and structurally characterized by \textit{Brauer} and \textit{Zintl} in 1937. \cite{brauer1937konstitution}. It crystallizes in the hexagonal space group \textit{P}6$_3$/\textit{mmc}. The structure can be regarded as an AB-stacking of 2D honeycomb layers composed of Na and Bi, where the Bi atoms are further coordinated along the \textit{c}$_{hex}$ axis by Na atoms that are located between the layers, resulting in a trigonal bipyramidal coordination sphere for Bi. 
This compound was predicted to be a 3D Dirac semimetal in 2012 by band structure calculations \cite{wang2012dirac}. A pair of bulk Dirac points that are located along the $\Gamma$-A line in the first BZ was subsequently confirmed by ARPES. A band inversion caused by SOC of Na-3\textit{s} and Bi-6\textit{p} states near the the $\Gamma$ point on this high symmetry line has been identified as the origin of the 3D Dirac state. \cite{liu2014discovery}

\begin{figure*}[htbp]
  \centering
  \includegraphics[width = \textwidth]{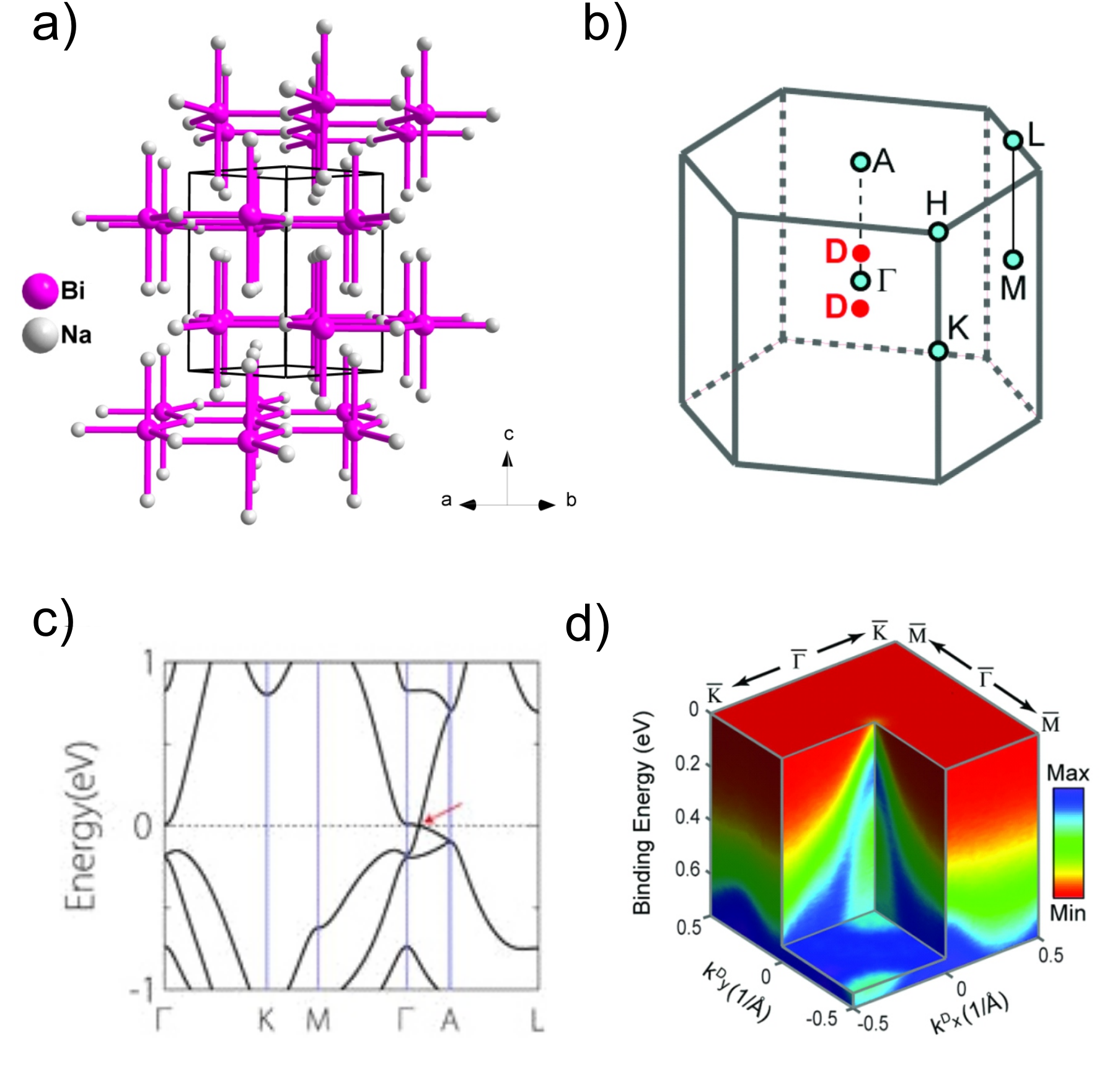}
  \caption{(a) Crystal structure of Na$_3$Bi, (b) first Brillouin zone with high symmetry points and highlighted Dirac points, adopted from \cite{liu2014discovery}, (c) bulk band structure, adopted from \cite{guan2017artificial}, (d) 3D intensity plot of the ARPES spectra at the Dirac point, adopted from \cite{liu2014discovery}}
  \label{fig:Fig5}
\end{figure*}

At about the same time, the low temperature polymorph of Cd$_3$As$_2$ was also confirmed to be a DSM by different groups by means of photoemission spectroscopy.\cite{borisenko2014experimental, liu2014stable, neupane2014observation} Cd$_3$As$_2$ exists in three different polymorphs. The structure of the low temperature (LT) polymorph can be described as a supercell of a distorted antifluorite type with ordered Cd vacancies. The space group of the low temperature form has been debated until 2014, even after the experimental confirmation of it being a Dirac semimetal. In 1964, the space group had been reported to be \textit{I}4$_1$/\textit{acd} (no. 142), but a proper determination of the atomic positions was missing. \cite{zdanowicz1964crystal} Four years later the space group was proposed to be \textit{I}4$_1$\textit{cd} (no. 110) on the basis of single crystal X-ray data.\cite{steigmann1968crystal} The lattice parameters were in line with the structure reported earlier, the only difference being a missing center of inversion in the latter. This center of inversion plays a crucial role with respect to the electronic structure - its absence would cause bands to be spin split. Indeed, a re-examination of the structure by \textit{Cava et al.} in 2014 \cite{ali2014crystal} confirmed the centrosymmetric space group \textit{I}4$_1$/\textit{acd}. Although the calculated band structure in the correct space group is similar to the previously reported one in the non-centroymmetric setting of the structure \cite{wang2013three}, there is an important difference: Spin-splitting is prohibited due to inversion symmetry, hence all bands are two-fold degenerate as in the 2D analog graphene. 
A further binary compound that was found to be a Dirac semimetal is PtSe$_2$, a compound that has been known since the 1920s. \cite{thomassen1929kristallstrukturen}. This layered material crystallizes in the CdI$_2$-type structure and was characterized as a type II DSM (i.e. the Dirac cone is tilted as in type II Weyl semimetals). \cite{zhang2017experimental}
Interestingly, not only compounds, but also certain allotropes of elements are potential materials with non-trivial band topology. In 2013 physicists from W{\"u}rzburg concluded from ARPES spectra that $\alpha$-Sn grown on InSb(001)-surfaces shows the characteristic band structure features of a strong TI, although there was no evidence for a band gap. \cite{barfuss2013elemental} Further investigations by another group on epitaxially produced thin films of $\alpha$-Sn on InSb(111) concluded this allotrope to be a 3D DSM from ARPES spectra, again as a result of lattice mismatch. \cite{xu2017elemental} Further measurements will be needed to resolve this discrepancy.



Early predictions claim Au$_2$Pb to turn from a DSM to a Z$_2$-topological metal at low temperatures. A Z$_2$-topological metal is a semimetal that has a continuous pseudo band gap in the vicinity of the Fermi level (but with bands crossing \textit{E}$_F$) and that hosts topological surface states. The DSM state is predicted to appear in the cubic high temperature phase which becomes stable above 100 K. Interestingly, the low temperature polymorph is superconducting at low temperatures (\textit{T}$_c$ = 1.18 K). \cite{schoop2015dirac} 
Electronic structure investigations on ternary 1:1:1 compounds of the ZrBeSi-type and the LiAlGe-type also identify candidates for Dirac semimetals such as BaAgBi, YbAuSb or LaAuSb, and further predictions on a variety of materials were made in \cite{gibson2015three,seibel2015gold}. However, none of these predictions has been experimentally verified as yet. 
Another general idea to realize a DSM are solid solutions of compounds that are trivial insulators at one end (\textit{x} = 0) and topological insulators on the other end (\textit{x} = 1). At a certain value of \textit{x}, a critical point with a DSM state should be reached. This seems difficult to realize, since a very precise control of the \textit{x}-parameter is necessary. However, two experimental observations of the DSM state were made on the systems TlBiS$_{2-x}$Se$_x$ (\textit{x} = 0.50) \cite{novak2015large} and Pb$_{1-x}$Sn$_x$Se (\textit{x} = 0.23) \cite{liang2013evidence}.
Another possible member of the family of 3D DSMs is the layered compound ZrTe$_5$. However, the question whether it is a TI or DSM is still discussed controversially. \cite{li2016experimental, chen2015optical, chen2015magnetoinfrared} Theory predicts that this material is very strain sensitive and undergoes a phase transition from a weak to a strong TI via a DSM state. \cite{fan2017transition} 
Finally, CuMnAs was the frist antiferromangetic Dirac semimetal predicted. In the antiferromnetic phase, while \textit{T}-symmetry and \textit{I}-symmetry are broken individually, the product of both still holds, which leads to four-fold degenerate Dirac cones in its antiferromentic state \cite{tang2016dirac}. Some early experimetal evidence for this proposal exits \cite{emmanouilidou2017magnetic}.

\subsection{Weyl semimetals}
\begin{figure*}[htbp]
  \centering
  \includegraphics[width = \textwidth]{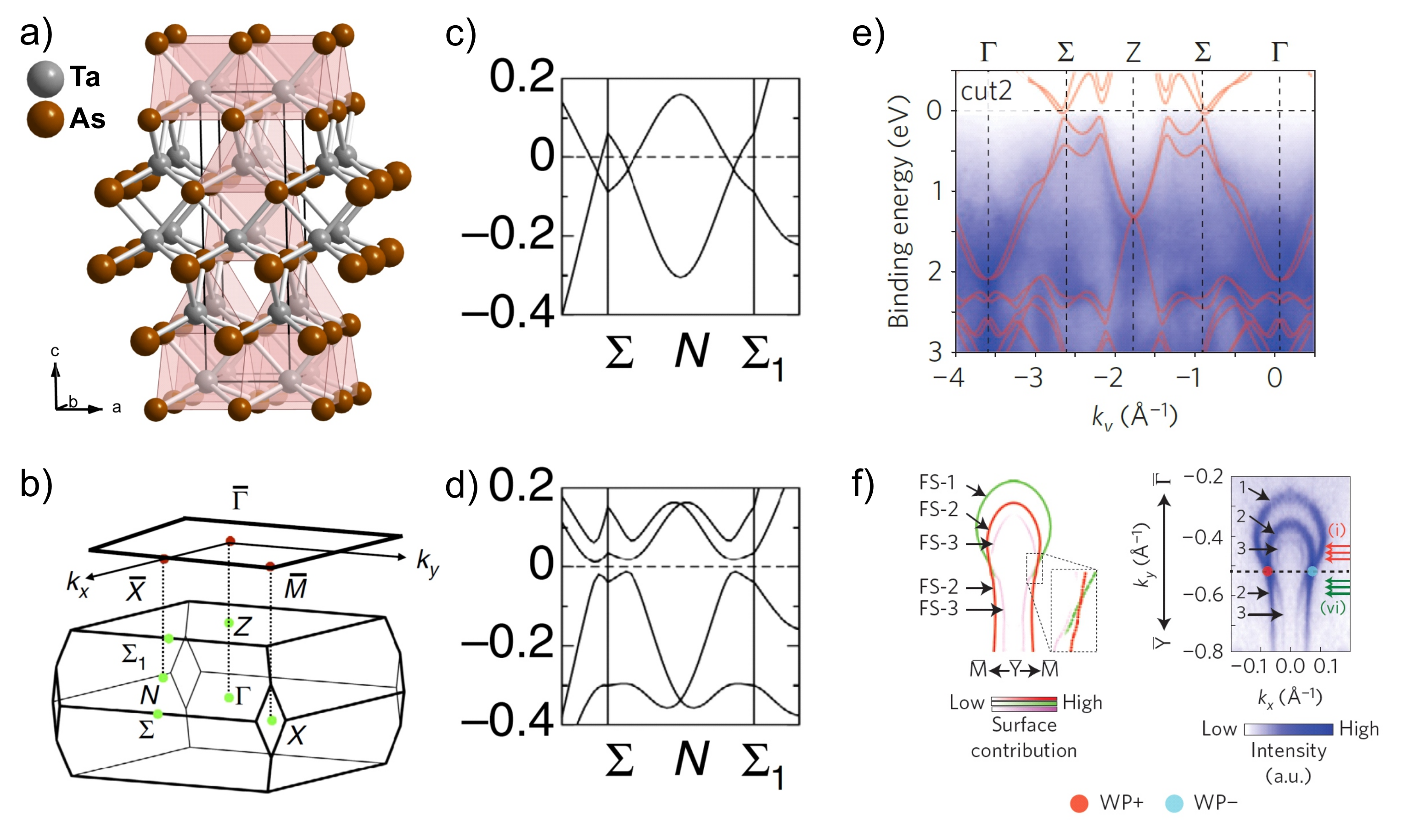}
  \caption{(a) Crystal strcuture of TaAs, (b) Brillouin zone, (c) band strcuture without and (d) with SOC, (e) photoemission spectrum with overlaid calculated band structure, (f) claculated and measured Fermi surface, dispayeing the Fermi arcs, which are the signature to identfy WSMs; figures (b), (c) and (d) adopted from \cite{huang2015weyl}; figures (e) and (f) adopted from \cite{yang2015weyl}}
  \label{fig:Fig7}
\end{figure*}

As outlined above, WSMs are harder to predict and identify via band structure calculations, since Weyl points are often not located a high symmetry lines or planes in the BZ. The first class of materials predicted to be WSMs were \textit{T}-breaking iridates with pyrochlore stucture, \textit{R}$_2$Ir$_2$O$_7$ (\textit{R} = Y,  Pr) \cite{wan2011topological} but an experimental confirmation is still missing. In 2015, the family of the isotypic monopnictides, TaAs, TaP, NbAs and NbP, was predicted and experimentally confirmed via ARPES to host type I Weyl crossings through \textit{I}-symmetry breaking. \cite{huang2015weyl, yang2015weyl, lv2015experimental, xu2015discovery, xu2015observation}  
Their crystal structure was first determined by powder \cite{boller1963transposition} and later by single crystal X-ray diffraction \cite{murray1976phase}. They crystallize in the non-centrosymmetric space group \textit{I}4$_1$\textit{md} (no. 109) with two crystallographically inequivalent positions. Both atom types have a six-fold trigonal prismatic coordination sphere that is face- and edge-linked in the (110) direction, resulting in 1D infinite edge- and face-linked prisms forming 2D layers. Along the \textit{c}-axis, these layers are rotated by 90 degrees and edge-linked to the layers below and above (see Fig. \ref{fig:Fig7}). 
A hint to the existence of Weyl points in these systems is given by a scalar-relativistic electronic structure calculation (where SOC is turned off): Without SOC, band crossings occur at high symmetry lines, forming a nodal line. When SOC is applied, the nodal line is gapped, and no crossing appears along the $\Gamma$-Z line where a \textit{C}$_4$ rotation axis protects the crossings in the presence of SOC.
Due to the lack of \textit{I}-symmetry, SOC also induces a spin-splitting and thus the nodal line is not fully gapped, 
but some Weyl crossings arise away from high symmetry lines. 
The first ternary compound confirmed to be a magnetic \textit{T}-breaking WSM by \textit{Cava et al.} was YbMnBi$_2$. \cite{borisenko2015time} Sr$_{1-y}$Mn$_{1-z}$Sb$_2$ (y, z $<$ 0.1) is structurally similar to YbMnBi$_2$ and a further magnetic compound characterized as \textit{T}-symmetry breaking WSM. \cite{liu2017magnetic} 
Anthorer \textit{T}-breaking Weyl semimetal is the half-Heusler compound GdPtBi. Weyl nodes are observable in this material in the presence of a magnetic field. This compound further shows evidence for the chiral anomaly. \cite{hirschberger2016chiral}

Even the popular binary transition metal dichalcogenides host WSMs: WTe$_2$, the material with the highest non-saturating  magnetoresistance \cite{ali2014large}, as well as MoTe$_2$, a superconductor with pressure-dependent \textit{T}$_c$ \cite{qi2016superconductivity}, are the most prominent examples that were predicted to be type II WSMs. \cite{soluyanov2015type, sun2015prediction, wang2016mote} MoTe$_2$ is found to crystallize in three polymorphic structures: the hexagonal 2H ($\alpha$), the monoclinic 1T' ($\beta$), and the  orthorhombic Td ($\gamma$) modifications \cite{clarke1978low}, \cite{manolikas1979electron}, whereas for WTe$_2$, only the orthorhombic form is reported. \cite{yanaki1973preparation} From band structure calculations different electronic ground states are obtained for the three MoTe$_2$ polymorphs. The 2H-phase is a trivial semiconductor with a band gap of 0.81 eV, the 1T'-phase is classified as a TI, although the ground state is found to be semimetallic due to a pseudo gap with bands crossing \textit{E}$_F$ and the Td phase is predicted to be a WSM analogous to WTe$_2$.\cite{sun2015prediction, sankar2017polymorphic} 
The orthorhombic phases of both were investigated by ARPES to reveal the Fermi arcs and based on the predictions classified as type II WSMs.\cite{feng2016spin, deng2016experimental, huang2016spectroscopic}

Some ambiguity regarding these results should be mentioned, however. \textit{Balicas et al.} measured the Fermi surface of MoTe$_2$ with quantum oscillations 
and found that the Fermi surface is drastically different to what was predicted from DFT. According to their analysis, no Weyl nodes can be found in orthorhombic MoTe$_2$. \cite{rhodes2017bulk}
Since quantum oscillations measure the Fermi surface more accurately than ARPES, these results are significant and should not be disregarded. The two-dimensional monolayers of these materials, which are also of interest in this context, are discussed in the "2D Materials" part below.
LaAlGe, another ternary compound that crystallizes in the LaPtSi type structure was recently confirmed to be a type II WSM. In this case, the photoemission data was interpreted without the help of bandstructure calculations.\cite{xu2017discovery}  Further, TaIrTe$_4$ \cite{koepernik2016tairte} was verified to be a type II WSM by ARPES. \cite{haubold2017experimental}  \\


\subsection{Nodal line semimetals}

Although there are numerous predictions for NLSMs (e.g. the antiperovskite Cu$_3$PdN \cite{yu2015topological}, Ca$_3$P$_2$ \cite{xie2015new}, CaP$_3$ \cite{xu2017topological}, $\beta$-PbO$_2$ \cite{wang2017new}, CaAg\textit{Pn} (\textit{Pn} = P, As) \cite{yamakage2015line}, TiB$_2$ and ZrB$_2$ \cite{feng2017topological}, alkaline earth metals (Ca, Sr and Yb) \cite{hirayama2017topological}, SrSi$_2$ \cite{huang2016new}), there are only very few experimentally verified ones:  PbTaSe$_2$ \cite{bian2016topological}, PtSn$_4$ \cite{wu2016dirac} and ZrSiS as well as the isostructural compounds ZrSiSe, ZrSiTe, HfSiS and ZrSnTe  \cite{schoop2015dirac,neupane2016observation,hu2016evidence,lou2016emergence,takane2016dirac,topp2016non,chen2017dirac}. 
The first synthesis and structural characterization of PbTaSe$_2$ was carried out by \textit{Eppinga et al}.\cite{eppinga1980generalized} It crystallizes in the non-centrosymmetric space group $P\bar{6}m2$ (no. 187).  The structure is derived from 1H-TaSe$_2$ by intercalating Pb in the van der Waals gap. PbTaSe$_2$ is a Weyl nodal line semimetal, where SOC does not gap the line node. 
In addition, PbTaSe$_2$ is superconducting below 3.8 K \cite{ali2014noncentrosymmetric} and the line node has been visualized with ARPES \cite{bian2016topological}. However, PbTaSe$_2$ is not a pure NLSM, i.e. it has trivial bands interfering at the Fermi level, and so far, no exciting transport properties have been reported, presumably for that reason. 

The compound PtSn$_4$ was first synthesized and structurally characterized by \textit{Schubert et al.}\cite{schubert1950kristallstruktur}. However, a redetermination of the crystal structure revised the space group from \textit{Aba}2 (no. 41) to \textit{Ccca} (no. 68). \cite{kunnen2000structure} PtSn$_4$ possesses a layered structure of edge-sharing PtSn$_8$ quadratic antiprisms, with an AB stacking of the layers. This intermetallic compound has been discovered to be a NLSM due to its unusual transport properties that motivated the study of its electronic structure \cite{wu2016dirac}. Note, however, that PtSn$_4$ is a rather unusual type of NLSM, as it has an arc rather than a loop of Dirac states. 

Finally, ZrSiS and related compounds have a diamond shaped line node at the Fermi level and there are no trivial bands interfering, which is why these materials show the expected transport behavior (see section "Transport" below).\cite{xu2015two,schoop2015dirac,lv2016extremely,ali2016butterfly,singha2017large,pezzini2017unconventional}. 
The family of ZrSiS and related compounds crystallizes in the PbFCl-strcuture type, space group \textit{P}4/\textit{nmm} (no. 129), known from the mineral Matlockite. A square net of Si atoms is located in the \textit{ab}-plane and two ZrS-layers composed of quadratic ZrS$_4$ pyramids are sandwiched between the Si nets. 
In ZrSiS type materials, the line node at the Fermi level is gapped by SOC, which is why the lighter ones are the "better" NLSMs. \cite{schoop2015dirac} 

We should mention that CaAgAs and CaAgP have been predicted to be NLSMs as well \cite{yamakage2015line}, and although the electronic structure has not yet been confirmed, some typical transport phenomena have been observed in CaAgAs \cite{takane2017observation}. NLSMs have recently been reviewed by \textit{Ali et al.} \cite{yang2017symmetry} and thus we refer the reader to this reference for further detailed descriptions of the crystal and electronic structures of these materials.


\subsection{Nonsymmorphic topological semimetals}

Although nonsymmorphic symmetry opens exciting opportunities, the experimental evidence for these types of semimetals is currently limited to a very few cases. This is mostly related to the challenge of having the Fermi level at the crossing point, as discussed above. ZrSiS has been the first material where a crossing point that is a result of nonsymmorphic symmetry has been imaged with APRES \cite{schoop2015dirac}. This crossing however is below the Fermi level, at the X point.  In space group 129, the one ZrSiS is adopting, a fourfold degeneracy at the X, M, R and A points of the BZ (note that the R and A point are above X and M, respectively) is enforced by symmetry \cite{topp2017surface}. It has been shown with APRES that along the X-R (and M-A) line (along $k_z$) the bands remain four-fold degenerate. Further, it has been proposed that in a monolayer, a perfect Dirac cone protected by nonsymmorphic symmetry could be obtained, since there is no $k_z$ dispersion \cite{guan2017two,xu2015two}  These materials are therefore potential candidates for 2D Dirac semimetals, where SOC does not gap the crossing. What is problematic though is that within one layer, the bands are four-fold degenerate along the XM line without SOC. SOC can be used to lift the degeneracy along this line, while it will not affect it at the high symmetry points. This effect has experimentally been seen in isostructural and heavy CeSbTe \cite{topp2017effect}. 

\begin{figure*}[htbp]
  \centering
  \includegraphics[width = \textwidth]{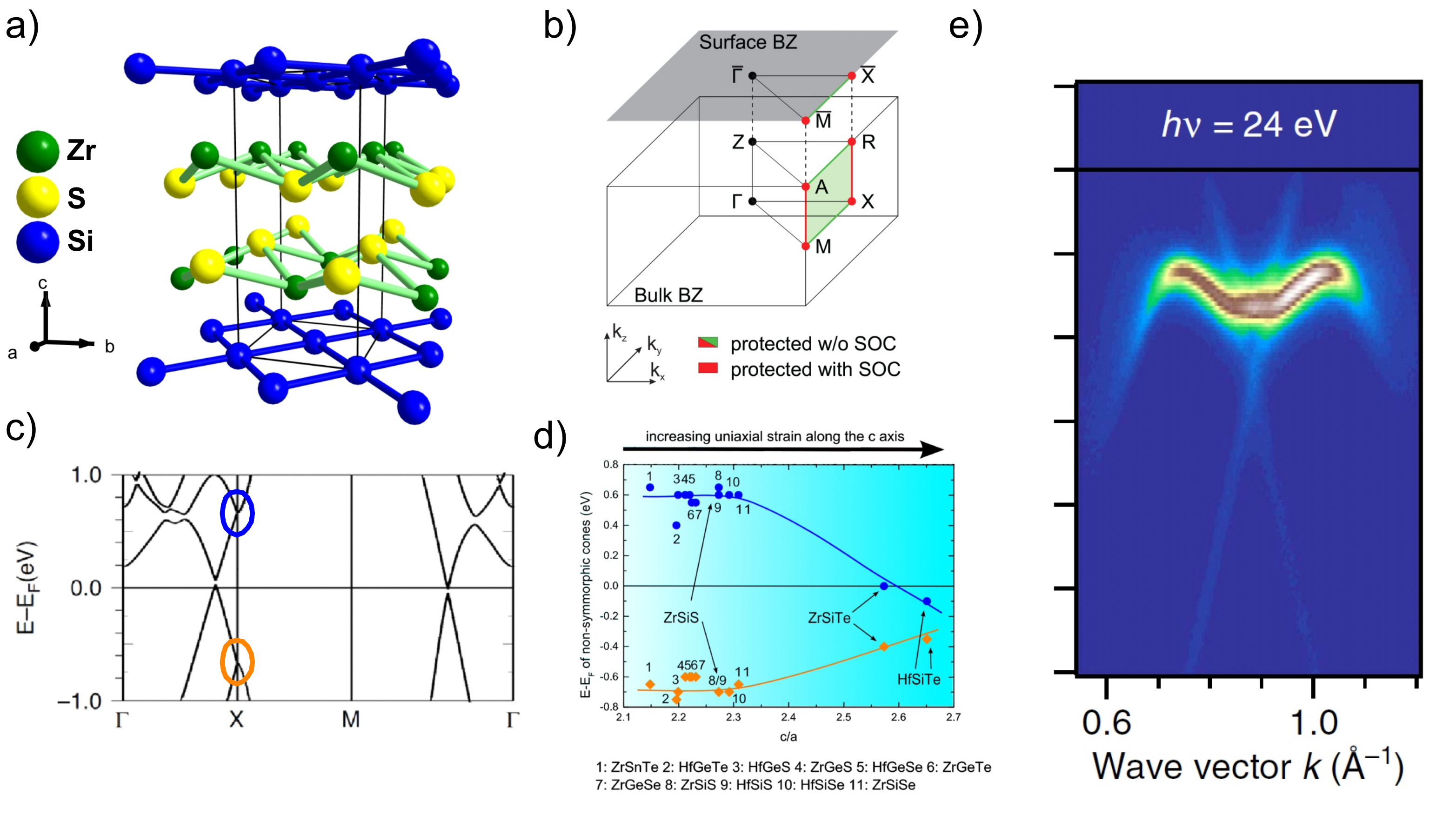}
  \caption{(a) Crystal structure of ZrSiS, (b) Brillouin zone in space group 129, with highlighted degeneracy enforced by nonsymmorphic symmetry (adapted from \cite{topp2017effect}), (c) bulk band structure of ZrSiS (figure adapted from \cite{schoop2015dirac}), the two degeneracies enforced by nonsymmorphic symmetry at the X point are highlighted in blue (above \textit{E}$_F$) and orange (below \textit{E}$_F$). (d) Effect of the \textit{c}/\textit{a} ratio of isostructural and isoelectronic analogs of ZrSiS on the nonsymmorphically induced degeneracies at X. While most compounds exhibit these crossings below and above the Fermi level, there are two exceptions, HfSiTe and ZrSiTe (adapted from \cite{topp2016non}). (e) ARPES spectrum of ZrSiS near X along $\Gamma$-X. Two bands cross at X due to the nonsymmorphic symmetry. Above the crossing, a very intensive surface state\cite{topp2017surface} is visible  (figure adapted from \cite{schoop2015dirac}).}
  \label{fig:Fig8}
\end{figure*}

 Further,  it is possible to use chemical pressure to shift the nonsymmorphic degeneracy to the Fermi level, which has been evidenced in ZrSiTe, where the larger Te atoms cause the \textit{c}/\textit{a} ratio to increase. The increase in \textit{c}/\textit{a} ratio shifts the bands in such a way the the four-fold degeneracy at X happens to be exactly at the Fermi level in ZrSiTe (Fig.\,\ref{fig:Fig8}d) \cite{topp2016non}. This way, a half-filled band is not necessary, but the crossing is not the only state at E$_F$, a few pockets interfere. Nonetheless, chemical pressure might be a useful tuning knob to realize exciting predictions in topological materials, such as new fermions in nonsymmorphic materials. 
 
 Lastly, as mentioned above, the effect of magnetism on nonsymmorphically protected band structures has been studied in CeSbTe. This compound, which is also isotopic with ZrSiS, features antiferromagnetic and fully polarized ferromagnetic-like 
 phases and thus several magnetic groups can be accessed. Depending on the magnetic group, the band degeneracy can be split or merged, resulting in many changes within the topological properties. It has been shown that magnetism can also result in new fermions. In CeSbTe both three-fold and eight-fold degenerate states can be achieved by introducing magnetic order \cite{schoop2017tunable}.
 
Besides the ZrSiS family, experiments have also been performed on SrIrO$_3$, which was predicted to have a nonsymmorphic line node. In the experiments it seems, however, as if there would be a trivial gap in reality.\cite{nie2015interplay} In addtion, InBi has been shown to host nonsymorphically proteded Dirac nodal lines that span a wide energy range along high symmtry lines. \cite{ekahana2017observation}


\subsection{2D Materials} \label{subsec:2D}

As we discussed in the first half of the review, graphene is the archeotypical topological semimetal, but it also linked to the prediction and discovery of topological insulators. There have been immense research efforts to make heavier versions of graphene to realize 2D TIs. Since this involves some interesting chemistry we briefly present an overview in this section. Further this section will discuss the advances on topological 2D materials that are structurally different from graphene.
The big hype on 2D materials triggered by graphene lead researchers to take a closer look at the heavier graphene analogs of group 14 elements. A nice review about the graphene homologs by \textit{Akinwande et. al} is recommend to the reader. \cite{molle2017buckled}  Indeed, the manufacturing of so-called 2D "\textit{X}enes" with \textit{X} = Si, Ge and Sn was proven to be possible in several cases. However, two major aspects make silicene, germanene and stanene differ from graphene: (i) It is so far not possible to obtain single free-standing layers of these materials without any substrate, and (ii) a buckling of the honeycomb layer is more stable than the ideal planar geometry. The buckling is a consequence of the larger bond lengths in silicene, germanene and stanene. 
Only if the sheet becomes buckled, the atoms get close enough such that the \textit{p}$_z$-orbitals that form the delocalized $\pi$-system still can overlap. 
Nevertheless, the formation of a delocalized $\pi$-system is disturbed, resulting in mixed \textit{sp$^3$}- and \textit{sp$^2$}-hybridization. The features of the electronic structure of the heavy \textit{X}enes are similar to graphene's. Linear dispersing bands around K and K' exist, and a small direct band gap is opened and increases with increasing atomic number. From first principles calculations the band gap of silicene, germanene and stanene is found to be non-trivial, i.e. these free-standing \textit{X}enes would be 2D TIs. However, the band gap of silicene is predicted to be extremely small (1.5 - 2.0 meV)\cite{liu2011quantum} and can therefore not be detected experimentally. 
As mentioned above, none of the heavier analogs of graphene were obtained as free-standing 2D monolayers yet. A lot of effort was spent on growing silicene, germanene and stanene on suitable substrates. However, the electronic properties of epitaxially produced \textit{X}enes differ from the predicted ones for the free-standing layer. The interaction with the substrate causes reconstructions and the degree of buckling varies in a broad range, depending on the kind of substrate. For example, from ARPES measurements it is evident that silicene grown on Ag(110) and Ag(111) surfaces shows linearly dispersing bands, but a trivial gap of approx. 0.6 eV is opened at K.\cite{de2010evidence, vogt2012silicene}
Thin film growth of \textit{X}enes thus offers many possibilities to manipulate the electronic structure of potential topologically non-trivial low dimensional systems.

\begin{figure*}[htbp]
  \centering
  \includegraphics[width = \textwidth]{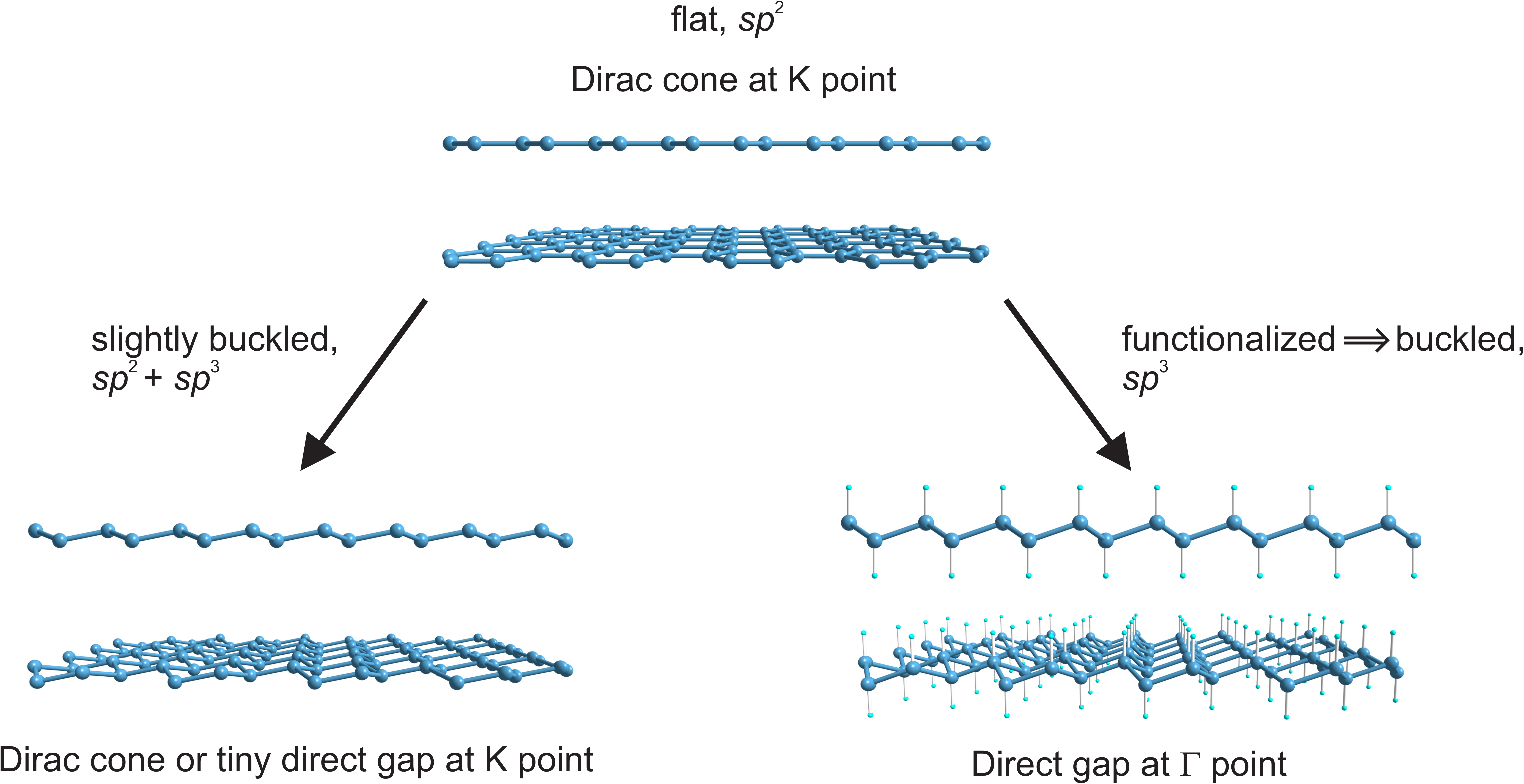}
  \caption{From flat \textit{sp$^2$} hybridization in graphene to slight buckling as found for the \textit{X}enes (\textit{X} = Si, Ge Sn), and functionalization to \textit{sp$^3$} hybridized \textit{X} in \textit{X}anes}
  \label{fig:Fig9}
\end{figure*}

A chemical way to stabilize 2D nanosheets of the group 14 elements was explored by \textit{Goldberger et al}.\cite{bianco2013stability} Starting from the Zintl-phase CaGe$_2$ which contains a 2D \textit{X}$^-$ polyanion that is isostructural to the layers in trigonal As, they were able to synthesize the so-called "\textit{X}anes" \textit{XR} (\textit{X} = Si, Ge, Sn; R = H or organic ligand) - covalently functionalized, buckled \textit{X} sheets with \textit{sp$^3$}-bonded \textit{X}-atoms.\cite{jiang2016tailoring} Due to \textit{sp$^3$}-hybridization, a trivial global band gap is opened, and a direct transition at the $\Gamma$ point is observed. 
A possible way to drive \textit{X}anes into topological insulators is predicted by \textit{Li et al.}\cite{si2014functionalized}. They show with full-relativistic DFT calculations (SOC included) that iodine functionalization of a Ge layer causes an inversion of bands at $\Gamma$. The origin of the non-trivial topology is therefore not the delocalized $\pi$-system as in graphene, but inverted $\sigma$-bonds.
Thus, there is a possibility to create TIs based on the buckled versions of the heavier analogs of graphene, but the band inversion will be based on $\sigma$-bonds, rather than $\pi$-bonds, and in that sense these materials are somewhat more similar to HgTe than to graphene.

Going from group 14 to group 15 elements, the orthorhombic allotrope of phosphorus, also known als black phosphorus, has been intensely investigated during the last few years, mainly due to the fact that it is the precursor to phosphorene, the exfoliated 2D version of the layered bulk material. It is thus not surprising that this material has been heavily explored in the context of topological semimetals. The electronic structure of few layer \textit{o}-P can be manipulated by K-doping, thus leading to a tunable band gap, caused by the giant Stark effect.\cite{kim2015observation} At high doping concentrations, a semiconductor to semimetal transition occurs, the band dispersion in \textit{k}$_x$ becomes linear and a Dirac crossing that classifies this system as a DSM is revealed by ARPES. 

The group 6 TMDs MoTe$_2$ and WTe$_2$ that were already mentioned above as bulk materials show different electronic characteristics as monolayers. The semiconducting state of 2D-1T'-TMDs was earlier predicted to be non-trivial. \cite{qian2014quantum} From the 2H-bulk phase to 1T'-MoTe$_2$, the monoclinic high temperature phase, the structural change is explained by a Peierls distortion of the Mo atoms, but the Peierls distortion does not open a band gap. The band gap opening is a quantum confinement effect, i.e. a result of reducing the bulk phase to a few layer 2D system, as described by band structure calculations and revealed by transport measurements that show a metal to semiconductor transition. \cite{keum2015bandgap} In monolayer WTe$_2$, the predictions about the 2D ground state have been probed by means of transport measurements. The interior does indeed become insulating and metallic conductivity of the edge states is observed in a monolayer, but absent in a bilayer. \cite{fei2017edge} This is a strong evidence for 2D-1T'-WTe$_2$ being a 2D TI.

\section{Properties and Applications of Topological Semimetals}

\subsection{Surface States}

The prominent electronic structure features of TSMs are in the bulk - unlike in TIs, where the surface states are the feature of interest. However, surface states can also play an important role in TSMs. Most prominent are probably the so-called Fermi arcs that appear on surfaces of Weyl semimetals. These arise due to the different chirality of the Weyl monopoles; the two monopoles will be connected by a surface state \cite{wan2011topological}. This surface state will be an open arc rather than a closed loop as is usually the case for surface states. This open orbit cannot exits by itself, which is why the electrons in such a Fermi arc surface state will enter the bulk to connect to another Fermi arc on the bottom surface\cite{wan2011topological}. The two arcs on opposite surfaces will form a loop to complete the path for the electrons. This path has been experimentally seen to exist: \textit{Analytis et. al.} took as sample the DSM Cd$_3$As$_2$ (note that Dirac semimetals become Weyl semimetals if a magnetic field is applied) and used a focused ion beam for sample thinning such that the surface states dominate the transport\cite{moll2016transport}. From the resistivity measured under an applied field the authors were able to extract a signature that corresponds to electrons moving along the Fermi arcs and shuttling through the bulk to reach the corresponding arc on the other surface.

Nonsymmorphic materials can also show surface states, either as semimetals or insulators. If the bulk is insulating, 
in rare cases the bulk bands can be connected in a special way that forces surface states to appear, which are topological. \cite{wieder2017wallpaper} There are a few cases of such nonsymmorphic topological insulators, but only one, exhibiting the "hourglass" surfaces state  \cite{wang2016hourglass} (called this way because it is shaped like an hourglass), has been seen experimentally \cite{ma2017experimental}. Semimetals that are nonsymmorphic can also show surface states: ZrSiS and related materials all show very prominent surface states in their ARPES spectra \cite{schoop2015dirac,takane2016dirac,lou2016emergence,topp2016non,neupane2016observation,hosen2017tunability}. These arise if a nonsymmorphic symmetry element is broken at the surface and thus the degeneracies in the bulk are no longer enforced at the surface. This results in 2D "floating" bands that exist only at the very top layer of the surface. While these states are topologically trivial, they are different from other surface states that appear commonly, such as Shockley states or quantum well states \cite{topp2017surface}.

\subsection{Transport}

A lot of attention is given to the electrical transport properties of topological semimetals. Soon after the confirmation that Cd$_3$As$_2$ is indeed a Dirac semimetal, \textit{Ong et al}. reported ultrahigh electron mobility in samples that were grown out of a Cd flux \cite{liang2015ultrahigh}. The mobility reaches nearly 10$^7$ cm$^2$ V$^{−1}$ s$^{−1}$ at 5 K, which is about 2 orders of magnitude higher than what is found in a free standing monolayer of graphene \cite{bolotin2008ultrahigh}. They also found evidence that this high mobility arises from a protection from back scattering. After this study, similar high mobilities have been found in many more Dirac and Weyl semimetals (see e.g. \cite{shekhar2015extremely,ali2016butterfly,ali2015correlation}) and it seems now to be a commonly observed property in TSMs. High mobility usually also causes high transversal magnetoresistance, since these two properties are connected via $MR$ (ratio) = $1 + \mu_{av}B^2$ \cite{ali2015correlation}.  Thus, high and unsaturated transversal magnetoresistance is another common property of topological semimetals. \cite{ali2014large,du2015unsaturated} The high mobility is also the reason why quantum oscillations already appear at small fields in topological semimetals. While such oscillations appear in any material that has a Fermi surface 
 at low temperatures and very high fields, materials with low electron masses and consequently high mobilities show these oscillations at fields that are accessible in a common laboratory and they can persist to temperatures of nearly 100 K \cite{pezzini2017unconventional}. 

The sample quality still matters, however. While the common features are often even observed in lower quality samples, which suggests that they are indeed induced by the special electronic structure, the values of carrier mobility can be increased by several orders of magnitude if the crystal quality is improved \cite{ali2015correlation}.  In the case of Na$_3$Bi, for example, quantum oscillations are only visible in high quality crystals grown out of a Na-flux. \cite{kushwaha2015bulk}

Another prominent transport feature of TSMs is the chiral anomaly. This feature should arise in Weyl semimetals (or Dirac semimetals under an applied field) when a magnetic field is applied parallel to the current direction \cite{son2013chiral}. In this case, it was predicted that the Weyl fermions will switch their chirality, which will lead to a negative magnetoresistance. This effect has been observed first in Na$_3$Bi \cite{xiong_evidence_2015} and later in many more Dirac and Weyl semimetals \cite{huang2015observation,li2015negative,li2014observation,yang2015chiral}. However, doubts have been cast on the analysis in certain cases, since more effects can lead to a negative magnetoresistance and very careful alignment of contacts is crucial and further tests have to be performed to be certain that the observed effect really stems from the chiral anomaly \cite{Shekhar2015large,hirschberger2016chiral}.

In the nodal line semimetal ZrSiS, besides high mobility and magentoresitance, an enahced correlation effect under high applied magnetic fields was observed. \cite{pezzini2017unconventional} This observed, apparent gain of electron mass under the influence of a magnetic field has yet be understood from a theoretical perspective.


\subsection{Optical properties}

The linear band dispersion in TSMs also effects their optoelectronic properties. Optical conductivity measurements can be used to distinguish between conventional, 3D and 2D Dirac electrons \cite{bacsi2013low}. 2D Dirac electrons exhibit a frequency independent optical conductivity, which was shown for graphene \cite{mak2008measurement}. In 3D Dirac (or Weyl) systems the optical conductivity changes linearly with the frequency, which has been shown for TaAs \cite{xu2016optical}, Cd$_3$As$_2$ \cite{neubauer2016interband} and ZrTe$_5$ \cite{chen2015optical}. The nodal line semimetal ZrSiS, while formally being a 3D system, has been shown to exhibit a frequency independent conductivity \cite{schilling2017flat}. Since a nodal line is a 2D object, this behavior is expected and thus it can be concluded that nodal line fermions behave more like graphene than 3D DSMs. Trivial materials do not exhibit such simple types of "power-law" optical conductivities and more complicated terms for their description are necessary.
 
As mentioned above, linear band dispersions have the potential to absorb photons with arbitrarily long wavelength. Other desirable features for next-generation optoelectronic devices such as infrared photodetectors \cite{chan2017photocurrents}, especially those used in high-speed optical communication systems, are fast carrier transport and short lifetimes of the photogenerated charge carriers. These result in short intrinsic response times of the device, thus allowing for operation at high frequencies.

 However, it is not possible to produce a photocurrent from just exposure to light in an ideal Dirac spectrum. As in trivial materials, the photocurrent vanishes without an external electric field. 
This is the case for ideal 2D Dirac semimetals because excitations are symmetric around the Dirac cone. Since \textit{I}-symmetry can be broken and a Weyl cone can be tilted in 3D, with the result of asymmetric excitations around the cone, the situation is different in type II Weyl semimetals. A large photocurrent proportional to the absorbed radiation is predicted to be stable up to room temperature - without applying a potential. In order for this scenario to become reality, we would need type II Weyl semimetals that are "clean", i.e. no other bands are interfering with the tilted Weyl cone at the Fermi level. Such materials have not yet been discovered.

The quality of sensing and detecting devices relies on the intrinsic optical properties of the applied materials and a steady optimization process of their performance is relevant for technological applications.  
The potential of the 2D analog of Cd$_3$As$_2$, graphene, as a high-performance photodetector was intensively investigated with respect to 
operation speed and broadband response.\cite{xia2009ultrafast,mueller2010graphene,gan2013chip,liu2014graphene,zhang2013broadband,pospischil2013cmos} 
In this context, first steps have been made to create ultra-fast broadband photo-detectors as well as mid-infrared optical switches based on Cd$_3$As$_2$. \cite{wang2017ultrafast,zhu2017robust,yavarishad2017room} Cd$_3$As$_2$ hosts all necessary promising properties that already make graphene an attractive photosensitive material, while it offers the further huge advantage that it interacts with light way better due to its 3D nature. 
Photodetectors based on a Au-Cd$_3$As$_2$-Au architecture exhibit an ultrafast photocurrent response of 6.9 ps, which is one order of magnitude faster than commercially available state-of-the-art systems. 
This extremely short response time is attributed to the very fast relaxation time of photoexcited carriers in Cd$_3$As$_2$. Rapid photo relaxation has been seen in all kinds of different types of Dirac semimetals. \cite{Weber2017similar} The future for applications of TSMs in optical devices thus seems bright.

\subsection{Catalysis}

Topology can have implications far beyond the transport properties of materials. This is particularly evident for chemistry happening at the surface of a material, as it is the exotic surface states that are the hallmarks of the non-trivial topology of the bulk bands of TIs, WSMs, and possibly also DSMs. Depending on the nature of the system, topological surface states (TSSs) are protected by the bulk crystal symmetry (time-reversal, nonsymmorphic etc) and are thus robust against backscattering and localization effects by (nonmagnetic) impurities, point defects and the like.  This feature distinguishes TSSs from trivial surface states, which are highly sensitive to the details of the surface and heavily influenced by defects or other kinds of disorder, or surface reconstruction. As a direct consequence, topological materials are set to provide a perfect platform for studying surface-related phenomena such as heterogonous catalysis, where the state of the surface is one of the most important, but also most difficult to describe parameters influencing catalytic activity. This is particularly true for catalysis studied in "realistic" environments, i.e. surfaces under ambient conditions or even surrounded by solvents, and thus far away from idealized UHV conditions.  
Bearing the unique robustness of TSSs in mind, let us consider the more specific influence of topology on catalysis. Take for example transition metal catalyzed surface reactions. Trends in catalytic activity are often rationalized by \textit{d}-band theory, which assumes that the nature of the metal - adsorbate interaction is determined by the hybridization of the metal \textit{d}-band with the bonding $\sigma$-orbital of the adsorbate. The occupation of the resulting bonding and antibonding metal - adsorbate \textit{d}-$\sigma$ states and thus the binding strength in turn is determined by the occupation and hence, energy (relative to the Fermi level) of the \textit{d}-band center. \cite{hammer1995electronic, hammer1995gold} This means that stronger metal - adsorbate bonding occurs for higher \textit{d}-band centers and weaker bonding for lower \textit{d}-band centers, which will have particularly pronounced effects if the rate determining step in heterogeneous catalytic reactions is given by the adsorption free energy of the gaseous or liquid reactant. This is the case for (electro)catalytic reactions following the Sabatier principle, for example hydrogen adsorption on metal electrodes in the hydrogen evolution reaction (HER). Plotting an observable for the catalytic activity, such as the exchange current density in electrocatalytic HER, against the adsorption free energy of hydrogen for different transition metals, will result in an activity scaling relationship known as "volcano plot", which can similarly be observed for a number of reactions involving oxygen and carbon-based adsorbates. Bearing this model in mind, it is intuitive that TSS can be an additional source of modifying catalyst - adsorbate interactions, as first pointed out by \textit{Chen et al.} \cite{chen2011co} In this work it was predicted by theoretical calculations that the TSS of Bi$_2$Se$_3$, covered by a thin layer of Au, would act on the adsorption energy of CO and O$_2$ on Au@Bi$_2$Se$_3$ in two different ways, depending on the relative energies of the surface Dirac cones and the antibonding 2$\pi$* orbitals of the adsorbates: While electrons are transferred from the antibonding 2$\pi$* orbital of CO to the TSS to strengthen the Au-CO bond, they are transferred from the TSS into the 2$\pi$* orbitals in the case of O$_2$ adsorption. Although the latter again enhances the Au-O$_2$ binding strength, the opposite direction of static electron transfer leads to more facile dissociative desorption of O$_2$ from the surface. It is obvious that such modifications of the binding energies will have pronounced effects on catalytic conversions in which surface adsorption and dissociation of reactants are rate-determining steps, like in CO methanation or water oxidation. Similar findings were reported by \textit{Yan et al}. who studied the interactions between different metal nanoclusters with the Bi$_2$Se$_3$ substrate, and between the clusters and oxygen atoms adsorbed to their surface via calculation of the adsorption free energy \textit{E}$_{ad}$(O). It is found that irrespective of the metal, the presence of TSS enhances the binding energy of oxygen in all cases, albeit to a different extent. As a consequence, TSSs can be considered an additional variable for tuning the metal - adsorbate interactions besides the inherent interaction strength given by \textit{d}-band theory. In the specific case investigated here, the catalytic activity towards oxidation is found to be enhanced through the TSS for Au, which shows an intrinsically weak Au-O binding strength, while the catalytic activity is significantly decreased for Pt and Pd, as the metal-oxygen binding becomes too strong in the presence of TSS, thus slowing down the desorption process of the products, which becomes rate limiting. Quite intuitively, this trend can be transferred to the hydrogen evolution reaction, which likewise follows the Sabatier principle and is thus sensitive to the intrinsic metal - adsorbate binding strength, modified by the presence of TSSs. \cite{xiao2015toward}
Overall, these predictions do not only underpin the possible role of TSSs as a tunable "electron bath" and additional parameter to adjust the catalyst - adsorbate interactions, but also highlight that - contrary to \textit{d}-band theory which implicitly assumes similar \textit{sp}-contributions for all transition metals - the \textit{sp}-band derived contributions of the TSS can have a significant - and differing - effect on the surface reactivity of transition metal catalysts, either through extrinsic (e.g. via the substrate) or through intrinsic contributions.\cite{chen2011co} 

\begin{figure*}[htbp]
  \centering
  \includegraphics[width = 0.7\textwidth]{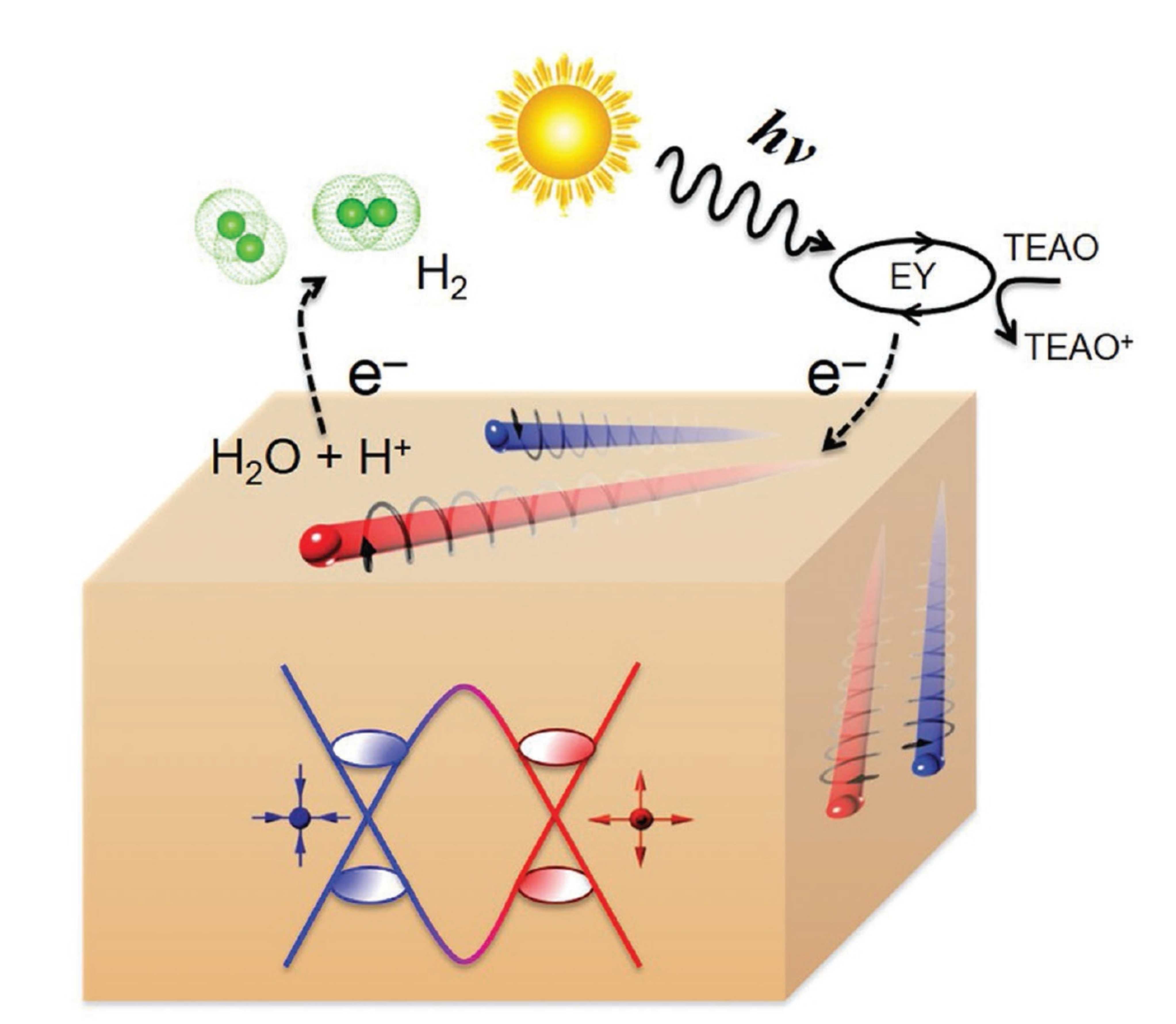}
  \caption{Schematic diagram of a topological Weyl semimetal for catalyzing the dye-sensitized hydrogen evolution, figure adopted from \cite{rajamathi2017weyl}}
  \label{fig:Fig11}
\end{figure*}

Looking at the electronic properties of topological materials, however, the impact of topology on catalysis is not limited to robust surface states and modified surface - adsorbate interactions. Intrinsic charge transport properties such as exceptionally high charge carrier mobilities associated with the linear dispersion around the Dirac points can play an additional role in enhancing the catalytic activity of topologically non-trivial materials. This can best be understood by considering electrocatalytic or photocatalytic reactions such as water splitting or CO$_2$ reduction where charge carriers of a semiconductor (i.e. the photoabsorber in photocatalysis) or a metal (electrocatalysis) have to be efficiently separated to prevent charge recombination, transported to the surface, and subsequently injected into a surface-bound substrate. Thus, the high conductivities as well as high charge carrier mobilities of DSMs and Weyl semimetals render them interesting candidates for catalytic conversions, as pointed out by \textit{Rao}, \textit{Felser} and co-workers who demonstrated photocatalytic activity for the hydrogen evolution reaction in the dye (Eosin)-sensitized Weyl semimetals 1T`-MoTe$_2$, NbP, TaP, NbAs, and TaAs.\cite{rajamathi2017weyl} While all systems were found to be active for HER, the trending activity, which shows a maximum for NbP producing 500 $\mu$mol g$^{-1}$ h$^{-1}$ of hydrogen, was rationalized by the differences in the Gibbs free energy of H adsorption at the surface of the catalysts, $\Delta$\textit{G}$_H$* (see \textit{d}-band model discussed above), which was found to be lowest for NbP. Although sacrificial agents are still needed to compensate for the lack of water oxidation activity of these systems, the authors point out that no signs of catalyst degradation is seen over the course of hours, thus boding well for the potential of DSM and WSM in (photo)electrocatalysis. Similar predictions can be made for TIs, while their bulk band gap may open additional perspectives for (non-sensitized) photocatalytic conversions.
An additional variable for increasing the activity, as the authors of the above study point out, is morphology: particle size reduction on going from single-crystalline to polycrystalline samples results in a twofold increase in activity. While this observation is not unexpected given the increased accessibility of active surface sites, it may be taken as a first hint that enhanced surface-to-volume ratios and higher defect levels do not adversely affect the surface protection of topological catalysts. Indeed, it has recently been shown that symmetry protected TSS are still present in TI nanoparticles where the Dirac cones at the surface are no longer continuous but break down into discrete states.\cite{siroki2017protection} The importance of this finding for the prospects of topological catalysis in the long run cannot be ranked highly enough, since catalytic activity, being a surface-related phenomenon, typically scales with the surface-to-volume ratio and, hence, size-reduction by nanostructuring is widely used as a convenient tool to further amplify catalytic activity. 
Transformations catalyzed by topologically non-trivial materials may be influenced by another degree of freedom inherent to TIs and WSMs, namely the chiral spin texture of TIs and WSMs, i.e. the fact that neither the TSS in TIs nor the chiral bulk states and surface Fermi arcs in WSMs are spin-degenerate.  This could have far-reaching implications - yet to be demonstrated - in catalytic conversions where the overpotential of certain reaction steps is sensitive to the spin state of the intermediates or products involved, such as in the water oxidation reaction. \cite{chretien20082, mondal2016spin, mtangi2015role} Spin-selective catalysis-driven by light, electrochemically or even thermally - is thus yet another field in which topological materials could make a difference by means of their inherent spin-polarization.

\section{Outlook}

The field of topological semimetals, while vastly growing, is still in its infancy.  However, new theoretical predictions of so far unknown phenomena and exciting physical properties in these materials are evolving rapidly. Materials that could be topological are also predicted immensely fast - but very often the experimental realization does not follow. Thus, the involvement of chemists who have a feeling for materials that are accessible and can be grown reliably is key to advance the field. In this spirit, this review serves as a guide for chemists and materials scientists to understand the basic principles of topology in general and the requirements topological semimetals should fulfill in particular. Armed with these insights, chemists should be able to provide a solid materials basis for this already burgeoning field with the ultimate goal that topology in materials science is made accessible to a broader audience and more chemists can join in. Finally, there are exciting possible applications for topological materials in different areas, while experimental evidence for them is still sparse. Clearly, in order to realize the full potential of topological materials, a larger library of confirmed compounds is of high demand and materials scientists are needed to unravel the vast promise topological matter has to offer in terms of new applications in quantum science, optoelectronics, catalysis and beyond.

\begin{acknowledgement}

We gratefully acknowledge the financial support by the Max Planck Society. The research at Princeton University was supported by a seed grant from the NSF-sponsored MRSEC at Princeton University, grant DMR-1420541. L.M.S. would like to thank Jennifer Cano, Barry Bradlyn, Lukas Muechler and Paul J. Chirik for helpful discussions.



\end{acknowledgement}





\providecommand{\latin}[1]{#1}
\makeatletter
\providecommand{\doi}
  {\begingroup\let\do\@makeother\dospecials
  \catcode`\{=1 \catcode`\}=2 \doi@aux}
\providecommand{\doi@aux}[1]{\endgroup\texttt{#1}}
\makeatother
\providecommand*\mcitethebibliography{\thebibliography}
\csname @ifundefined\endcsname{endmcitethebibliography}
  {\let\endmcitethebibliography\endthebibliography}{}

\end{document}